\pdfoutput=1
\documentclass[aps,prd,reprint,superscriptaddress,longbibliography,nofootinbib]{revtex4-2}
\usepackage{times}     
\usepackage{booktabs} 
\usepackage{graphicx}
\usepackage{amsmath}
\usepackage{amssymb}
\usepackage{amsfonts}
\usepackage{dcolumn}
\usepackage{dsfont}
\usepackage{latexsym}
\usepackage{rotating}
\usepackage{color}
\usepackage{latexsym}
\usepackage{bbm}
\usepackage{subfigure}
\usepackage{float}
\usepackage{epsfig}
\usepackage{natbib, hyperref}
\usepackage{verbatim}
\usepackage{bm}
\usepackage{amsthm}
\usepackage{eucal}
\usepackage{mathrsfs}
\usepackage{url}
\usepackage{braket}
\usepackage{ulem}
\usepackage{calligra}
\usepackage[T1]{fontenc}
\usepackage[utf8]{inputenc}
\usepackage{marvosym}
\usepackage[cal=cm]{mathalfa}
\usepackage{soul}
\usepackage{appendix}
\usepackage{tikz}
\usepackage{textcase}

\usepackage{color}

\usepackage{hyperref}
\hypersetup{
	colorlinks=true,
	final=true,
	linkcolor=blue,
	filecolor=magenta,
	urlcolor=blue,
	citecolor= blue,
	bookmarks=true,
}

\begin{document}
	\title{Symmetry-selective field-induced triplet superconductivity in Ising-superconductor monolayer NbSe\textsubscript{2}}
	
	\author{Sudipta Biswas}
	\email{sbiswas4396@iitkgp.ac.in}
	
	\author{Sudhansu S. Mandal}
	\email{sudhansu@phy.iitkgp.ac.in}
	
	\author{A. Taraphder}
	\email{arghya@phy.iitkgp.ac.in}
	
	\affiliation{Department of Physics, Indian Institute of Technology Kharagpur, Kharagpur 721302, West Bengal, India}

	\begin{abstract}
	   We investigate superconductivity in monolayer NbSe$_2$, an Ising superconductor, under an in-plane Zeeman field $h_x$, focusing on the emergence of symmetry selected equal-spin triplet pairing. Using a self-consistent Bogoliubov--de Gennes approach with realistic hopping parameters for monolayer NbSe$_2$ on a triangular lattice, we determine the energetically favored singlet pairing states in a range of chemical potentials $\mu$ 
	   for onsite, nearest-neighbor, and next-nearest-neighbor pairing channels, and map the resulting phase diagram in $(h_x,\mu)$ plane.
	   A momentum-resolved analysis reveals distinct dominant contributions in order parameters arise from the surroundings of $\Gamma$, $K$, and $K'$ points of the Brillouin zone.
	   We find that an in-plane magnetic field helps to induce a triplet pairing  whose symmetry is determined by the parent singlet state, upon opening the triplet interaction channel. For the nonlocal pairing channels, chiral-$d$ pairing is energetically favored and it induces  chiral-$p$ equal-spin triplet component with opposite chirality. We further find that Rashba spin-orbit coupling, relevant to substrate coupling and electrostatic gating, suppresses the superconducting orders and reduces the critical field. Our results establish a direct link between the symmetry of the parent singlet condensate and the emergent triplet superconductivity, highlighting monolayer NbSe$_2$ as a promising platform for field-tunable mixed-parity superconducting states.
	\end{abstract}
	\maketitle

	\section{\label{sec:I}Introduction}
	
	The realization of superconductivity in atomically thin transition-metal
	dichalcogenides (TMDs) has attracted considerable interest due to their
	distinctive electronic properties and the possibility of exploring
	superconductivity in the two-dimensional limit~\cite{Frindt1972,Costanzo2016,Tsen2016,Hsu2017,Ugeda2016,Baidya2021,Steinberg2025}.
	Monolayer NbSe$_2$ provides a particularly compelling platform for
	studying two-dimensional superconductivity and exhibits a robust
	superconducting state~\cite{Ugeda2016,Sohn2018,Khestanova2018,Xi2016}. Unlike its
	bulk counterpart~\cite{Foner1973,Noat2015}, monolayer NbSe$_2$ lacks inversion symmetry, leading to spin-split bands near the $K$ and $K'$ valleys due to strong
	spin-orbit coupling (SOC)~\cite{Xiao2012,Xi2016,Zhou2016,Wickramaratne2020}. Time-reversal
	symmetry locks the spin polarization at the $K$ and $K'$ valleys in
	opposite directions, while the $D_{3h}$ symmetry and preserved
	horizontal mirror symmetry constrain the spins to point predominantly
	out of the plane~\cite{Xiao2012,Xi2016,Zhou2016,Wickramaratne2021}.
	This spin-valley locking gives rise to Ising superconductivity and
	protects Cooper pairs against an in-plane magnetic field, allowing
	the upper critical field to substantially exceed the conventional
	Pauli limit~\cite{Lu2015,Saito2016,Xi2016,Zhou2016,deLaBarrera2018}.
	The robustness of Ising superconductivity against strong in-plane
	fields has subsequently been demonstrated in few-layer NbSe$_2$ and
	related systems~\cite{Xing2017,Sohn2018,Hamill2021,Samuely2021,Kuzmanovic2022,Samuely2023,Stefan2017},
	making NbSe$_2$ a promising platform for studying unconventional
	superconducting states under strong magnetic fields.
	
	The possibility of unconventional superconducting states in monolayer
	NbSe$_2$ provides a natural extension beyond its conventional
	phonon-mediated singlet $s$-wave state~\cite{Ugeda2016,Zheng2019,Shubham2024}.The combination of strong spin-orbit coupling, inversion-symmetry breaking, and valley degrees of freedom in the two-dimensional limit provides a natural setting for unconventional and topological superconductivity~\cite{Hsu2017,Moeckli2018,Sticlet2019,Shaffer2020,Akbar2024,Roy2024}. These unconventional pairing possibilities include superconducting states with nontrivial momentum-space structure arising from nonlocal pairing interactions~\cite{Pangburn2023}, while competing pairing channels have been proposed as a route to nodal and nematic superconductivity in NbSe$_2$~\cite{Cho2022,Hamill2021}.
	Collective-mode measurements have also provided evidence for a nearby
	triplet pairing instability~\cite{Wan2022}. These developments motivate
	the study of nonlocal singlet pairing beyond the conventional onsite
	state and raise the question of how the spatial structure of the pairing
	affects the superconducting response to an in-plane magnetic field~\cite{Pangburn2023}.
	
	The response to an in-plane field is particularly interesting in an
	Ising superconductor. The combined action of Ising spin-orbit coupling
	and Zeeman coupling can generate equal-spin triplet correlations,
	leading to field-induced mixed singlet-triplet superconductivity
	\cite{Moeckli2018,Moeckli2019,Moeckli2020,Wickramaratne2020,Tang2021,Kuzmanovic2022,Ilic2023}.
	For a conventional $s$-wave parent state, the induced triplet component
	has been shown to have $f$-wave character and to affect the upper
	critical field \cite{Moeckli2019,Haim2020,Moeckli2020}, while high-field
	tunneling experiments have provided evidence for triplet
	superconductivity in few-layers NbSe$_2$ \cite{Kuzmanovic2022}. This motivates us to study the thermodynamic stability of singlet pairing symmetries beyond onsite pairing channels and the corresponding emergence of triplet pairing symmetries upon application of in-plane magnetic field.
	
	
	In this paper, we investigate the superconducting state of monolayer NbSe$_2$
	under an in-plane Zeeman field using a self-consistent
	Bogoliubov--de Gennes (BdG) approach with a realistic single-band
	tight-binding model on the triangular Nb lattice~\cite{Bobkov_2024_A,Bobkov_2024_B}. We consider
	onsite, nearest-neighbor (NN), and next-nearest-neighbor (NNN)
	spin-singlet pairing channels and determine their thermodynamic
	stabilities around the reported chemical potential in NbSe$_2$. For the NN
	pairing sector, we compare the extended-$s$ and chiral-$d$ states
	through their condensation energies and obtain the corresponding
	$(h_x,\mu)$ phase diagram. We find that the chiral-$d$ state is
	favored near the relevant chemical potential. We further show that
	the NNN chiral-$d'$ state is likewise favored within the NNN pairing
	sector, but with a distinct momentum-space distribution: the NN
	chiral-$d$ state is predominantly supported by the $K/K'$ valleys,
	whereas the NNN chiral-$d'$ state is dominated by the $\Gamma$ pocket.
	Ater finding these self-consistent singlet states, we determine the
	symmetry of the field-induced equal-spin triplet component by solving
	the coupled singlet-triplet gap equations for all symmetry-allowed
	odd-parity channels. For the onsite $s$-wave state, the in-plane field
	induces an NN $f$-wave triplet component, whereas the chiral
	$d+id$ ($d-id$) state induces chiral $p-ip$ ($p+ip$) pairing with opposite chirality. For both
	NN and NNN chiral-$d$ parent states, the NN chiral-$p$ component is
	dominant, with a subdominant contribution from the corresponding NNN
	channel. The symmetry selection is governed by the
	$D_{3h}$ point-group structure of monolayer NbSe$_2$. Momentum-resolved
	calculations further reveal a contrasting momentum-space character
	between the parent singlet and induced triplet states: while the NN
	chiral-$d$ singlet is predominantly associated with the $K/K'$ valleys,
	the induced NN and NNN chiral-$p$ triplet component is concentrated around the
	$\Gamma$ pocket.
	We also examine the effects of a finite intrinsic
	triplet interaction and Rashba spin-orbit coupling, finding that the
	triplet interaction enhances the field robustness of the mixed-parity
	state, whereas Rashba coupling suppresses both singlet and induced
	triplet orders and reduces the critical field. These results establish
	a direct connection between pairing range, momentum-space structure,
	parent-state symmetry, and field-induced equal-spin triplet pairing in
	monolayer NbSe$_2$.

	The remainder of the paper is organized as follows. Section~\ref{sec:II} introduces the tight-binding model and Ising SOC for monolayer NbSe$_2$. Section~\ref{sec:III} presents the BdG formulation and relevant pairing channels. Section~\ref{sec:IV} presents our results for onsite, NN, and NNN pairing, including the field-induced triplet states and the effect of Rashba SOC. Section~\ref{sec:V} summarizes our results. Additional results are provided in Appendices~\ref{app:onsite_s_triplet_channels} and \ref{app:condensation_energy}, covering the onsite $s$-wave triplet channels and condensation-energy calculation, respectively, while the symmetry analysis of the field-induced triplet channels for the NN and NNN chiral-$d$ states are given in Appendices~\ref{app:chiral_d_triplet_channels} and \ref{app:NNN_chiral_d'_triplet_channels}, respectively.

	\section{\label{sec:II}Tight-binding model for monolayer $\mathrm{\mathbf{NbSe_2}}$}
	
\begin{figure}[t]
		\centering
		\includegraphics[width=\linewidth]{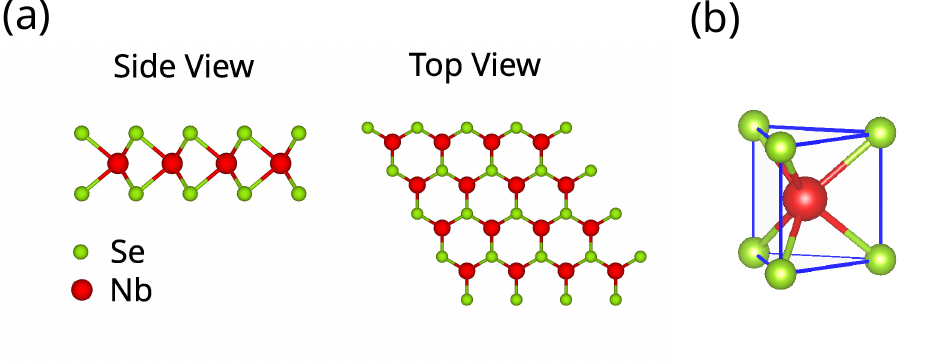}
		
		\caption{Crystal structure of monolayer $1H$-NbSe$_2$. 
			(a) Side and top views showing the triangular Nb layer sandwiched
			between two Se layers. The nearest-neighbor Nb--Nb distance defines
			the lattice constant $a$. (b) Trigonal-prismatic coordination of a Nb atom in the $1H$ phase~\cite{Liu2013}.}
		\label{fig:NbSe2_structure}
	\end{figure}

	We consider monolayer NbSe$_2$, shown schematically in Fig.~\ref{fig:NbSe2_structure}, modeled by a single-band tight-binding description on a triangular lattice that captures the low-energy physics near the Fermi level \cite{Bobkov_2024_B}. The Hamiltonian reads
	\begin{equation}\label{eq:normal_ham}
		H_0 =
		- \sum_{\langle ij \rangle_p,\sigma} t_p \, c_{i\sigma}^\dagger c_{j\sigma}
		- \mu \sum_{i,\sigma} c_{i\sigma}^\dagger c_{i\sigma}
		+ H_{\text{Ising}}.
	\end{equation}
	Here $c_{i,\sigma}$ is an electron annihilation operator at site $i$ in the plane of the Nb atoms of the NbSe$_2$ monolayer with spin $\sigma=\uparrow,\downarrow$, and $\mu$ is the chemical potential. We consider hopping processes up to the sixth nearest-neighbors \cite{Bobkov_2024_B} for describing the electronic structure of monolayer NbSe$_2$ as close as possible with minimum number of parameters. The hopping amplitudes $t_p$, where $p$ denotes the order of the nearest neighbor, are listed in Table~\ref{tab:tb_parameters}.
	
	\begin{table}[hbt]
		\caption{Parameters of the single-band tight-binding model for NbSe$_2$ taken from Ref.~\cite{Bobkov_2024_B}. All values of the hopping amplitudes and other energies are given in meV.}
		\label{tab:tb_parameters}
		\centering
		\begin{tabular}{ccccccc}
			\hline\hline
			\noalign{\vskip 1mm}
			$\mu$ & $t_1$ & $t_2$ & $t_3$ & $t_4$ & $t_5$ & $t_6$ \\
			\noalign{\vskip 1mm}
			\hline
			\noalign{\vskip 1mm}
			$-31.4$ & $-17.5$ & $-99.8$ & $7.8$ & $3.6$ & $14.3$ & $-0.5$ \\[1mm]
			\hline\hline
		\end{tabular}
	\end{table}
	
	Due to the lack of in-plane inversion symmetry together with the strong atomic spin-orbit interaction of Nb atoms, electrons experience an effective out-of-plane spin-orbit field. This gives rise to the characteristic Ising spin--orbit coupling, which locks electron spins along the $z$ direction with opposite orientations in the $K$ and $K'$ valleys. On the lattice, the Ising SOC can be incorporated as a spin-dependent nearest-neighbor hopping term~\cite{Cohen2024},
	\begin{equation}
		H_{\text{Ising}} =
		i \lambda_I
		\sum_{\langle ij \rangle_1}
		\nu_{ij}\, t_1
		\left(
		c_{i\uparrow}^\dagger c_{j\uparrow}
		-
		c_{i\downarrow}^\dagger c_{j\downarrow}
		\right),
	\end{equation}
	where $\nu_{ij}=\pm1$ depends on the bond orientation. The strength of the Ising SOC is controlled by the dimensionless parameter $\lambda_I$, which acts only on the first nearest-neighbor hopping.
	
	\begin{figure*}[t]
		\centering
		
		\begin{minipage}[t]{0.31\textwidth}
			\centering
			\includegraphics[width=\linewidth]{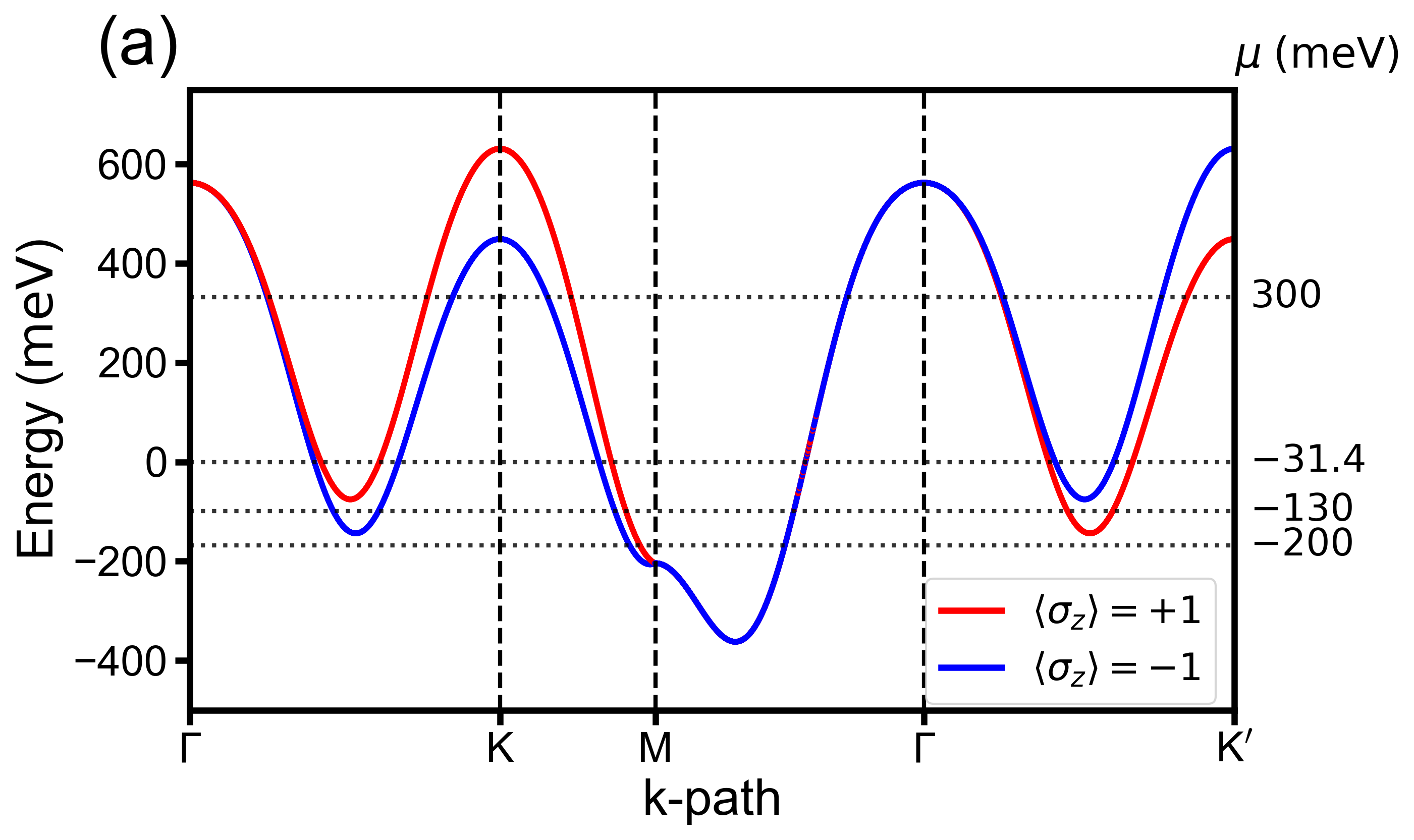}
		\end{minipage}
		\hfill
		\begin{minipage}[t]{0.68\textwidth}
			\centering
			\includegraphics[width=\linewidth]{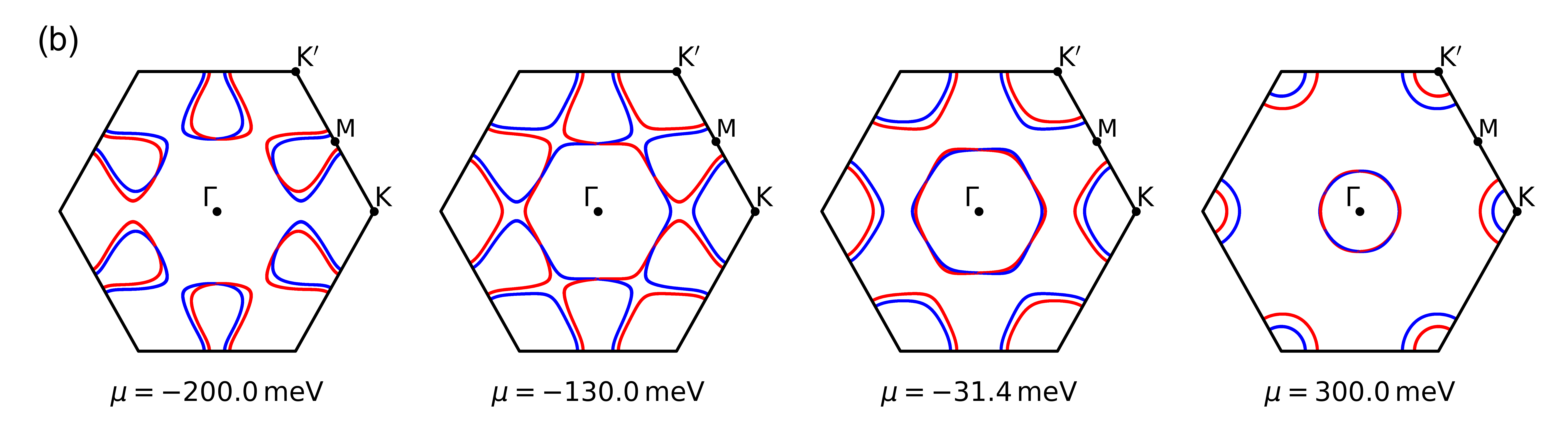}
		\end{minipage}
		\caption{(a) Spin-resolved normal-state band structure of monolayer NbSe$_2$ with Ising spin-orbit coupling ($\lambda_I=1.0$) along the $\Gamma$--$K$--$M$--$\Gamma$--$K'$ path. The color denotes the out-of-plane spin polarization $\langle\sigma_z\rangle$. (b) Spin-polarized Fermi surfaces for $\mu=-200$, $-130$, $-31.4$, and $300$~meV, showing the evolution of the Fermi-surface topology with chemical potential. Red and blue denote opposite out-of-plane spin polarizations.}
		\label{fig:normal_state}
	\end{figure*}
	
	By transforming the Hamiltonian $H_0$ into momentum space, we find
	\begin{equation}
	H_0	=\sum_{\mathbf{k}}
			\Psi_{\mathbf{k}}^\dagger
			\hat{H}_0(\mathbf{k})
			\Psi_{\mathbf{k}}
		=\sum_{\mathbf{k}}
		\Psi_{\mathbf{k}}^\dagger
		\left[
		\xi(\mathbf{k})\,\mathbb{I}
		+
		\xi_I(\mathbf{k})\,\sigma_z
		\right]
		\Psi_{\mathbf{k}},
		\label{total_Ising_Ham}
	\end{equation}
	where
	$\Psi_{\mathbf{k}}
	=
	(c_{\mathbf{k}\uparrow},\,c_{\mathbf{k}\downarrow})^{T}$
	is the spinor in spin space.
	The spin-independent dispersion $\xi(\mathbf k)$ includes hopping up to the sixth nearest neighbors \cite{Aikebaier_2022,Bobkov_2024_B}:
	\begin{align}
		\xi(\mathbf k)
		&= -2t_1[\cos(2\alpha)+2\cos\alpha\cos\beta]  \notag\\
		&\quad -2t_2[\cos(3\alpha-\beta)+\cos(2\beta)+\cos(3\alpha+\beta)] \notag\\
		&\quad -2t_3[\cos(4\alpha)+2\cos(2\alpha)\cos(2\beta)] \notag\\
		&\quad -2t_4(
		\cos(5\alpha+\beta)+\cos(5\alpha-\beta)
		+\cos(4\alpha+2\beta) \notag\\
		&\qquad +\cos(4\alpha-2\beta)+\cos(\alpha+3\beta)+\cos(\alpha-3\beta)) \notag\\
		&\quad -2t_5[\cos(6\alpha)+2\cos(3\alpha)\cos(3\beta)] \notag\\
		&\quad -2t_6[\cos(6\alpha-2\beta)+\cos(6\alpha+2\beta)+\cos(4\beta)]
		-\mu,
	\end{align}
	where the lattice constant is set to $a=1$, with
	$\alpha=k_x/2$ and $\beta=\sqrt{3}k_y/2$.
	The Ising SOC contribution is given by
	\begin{equation}
		\xi_I(\mathbf k)
		= 2\lambda_I t_1
		\left[\sin(2\alpha) - 2\sin\alpha\cos\beta \right],
	\end{equation}
	which is odd under $\mathbf{k}\rightarrow -\mathbf{k}$ and gives rise to the characteristic valley-contrasting spin splitting while preserving time-reversal symmetry.
	
	The Ising SOC lifts the spin degeneracy of the normal-state bands and generates opposite out-of-plane spin polarization in the $K$ and $K'$ valleys, as shown in Fig.~\ref{fig:normal_state}(a). For $\lambda_I=1.0$, the Ising spin splitting at the $K$ and $K'$ valleys, measured as half the difference between the spin-up and spin-down band energies, is approximately $91$ meV, while it vanishes at the $\Gamma$ and $M$ points. Consequently, the $K$ and $K'$ valleys host opposite out-of-plane spin polarizations, characteristic of spin-valley locking in monolayer NbSe$_2$.

	Figure~\ref{fig:normal_state}(b) shows the spin-polarized Fermi surfaces at several values of the chemical potential. As the chemical potential is varied, the Fermi-surface topology evolves, reflecting the underlying saddle-point structure and van Hove physics characteristic of monolayer NbSe$_2$~\cite{Kim2021}. In particular, the emergence of a $\Gamma$-centered pocket and persisting of the spin-polarized $K/K'$ valley pockets  signal a change in Fermi-surface topology, indicative of a Lifshitz transition. At higher chemical potentials, the Fermi-surface topology changes further as the pockets evolve toward the Brillouin-zone boundary. Throughout this evolution, the opposite spin polarizations of the $K$ and $K'$ valleys are preserved, reflecting the spin-valley locking induced by the Ising SOC.
	
	\section{\label{sec:III} Superconducting Hamiltonian and Pairing States in $\mathrm{\mathbf{NbSe_2}}$}
	
	Having established the normal-state electronic structure of monolayer
	NbSe$_2$, we now introduce superconductivity within the BdG formalism. In the presence of an external
	magnetic field, the normal-state Hamiltonian becomes
	\begin{equation}
		\hat{H}(\mathbf{k})
		=
		\hat{H}_0(\mathbf{k})
		+
		\mathbf{h}\cdot\bm{\sigma}
		=
		\xi(\mathbf{k})\,\mathbb{I}
		+
		\xi_I(\mathbf{k})\,\sigma_z
		+
		\mathbf{h}\cdot\bm{\sigma}\, .
	\end{equation}
	Throughout this work, an
	in-plane magnetic field is applied along the $x$ direction,
	$\mathbf{h}=(h_x,0,0)$, such that the Zeeman term reduces to
	$h_x\sigma_x$.
	
	\subsection*{Bogoliubov--de Gennes Formalism}
	
	Introducing Nambu spinor
	\(
	\Psi_{\mathbf{k}}
	=
	(c_{\mathbf{k}\uparrow},
	c_{\mathbf{k}\downarrow},
	c^\dagger_{-\mathbf{k}\uparrow},
	c^\dagger_{-\mathbf{k}\downarrow})^{T},
	\)
	on finds the BdG Hamiltonian in the form~\cite{Mohanta2014,Tang2021,Lie2026},
	\begin{equation}\label{eq:BdG_Hamiltonian}
		H_{\mathrm{BdG}}(\mathbf{k})=
		\begin{pmatrix}
			\hat{H}(\mathbf{k}) &
			\hat{\Delta}(\mathbf{k})\\
			\hat{\Delta}^{\dagger}(\mathbf{k}) &
			-\hat{H}^{T}(-\mathbf{k})
		\end{pmatrix}.
	\end{equation}
	The superconducting pairing matrix containing both spin-singlet and equal-spin triplet components is given by
	\begin{equation}\label{eq:pairing_matrix}
		\hat{\Delta}(\mathbf{k})
		=
		\Delta_s \phi_s(\mathbf{k})(i\sigma_y)
		+
		\begin{pmatrix}
			\Delta_{\uparrow\uparrow}\phi_t(\mathbf{k}) & 0 \\
			0 & \Delta_{\downarrow\downarrow}\phi_t(\mathbf{k})
		\end{pmatrix},
	\end{equation}
	where $\Delta_s$ denotes the spin-singlet pairing amplitude with even-parity form factor $\phi_s(\mathbf{k})$, while $\Delta_{\uparrow\uparrow}$ and $\Delta_{\downarrow\downarrow}$ represent equal-spin triplet pairing amplitudes associated with the odd-parity form factor $\phi_t(\mathbf{k})$.
	
	In this work, we consider spin-singlet onsite-$s$, extended-$s$, and chiral $d$-wave pairing states in the absence of magnetic field, and examine possible induction of  odd-parity equal-spin triplet chiral $p$-wave and $f$-wave pairing channels in presence of magnetic field. Different pairing channels are opened depending on onsite, NN, and NNN pairings. For bond-dependent pairing, the momentum-space form factor, in general, is given by
	\begin{equation}
		\phi(\mathbf{k})
		=
		\sum_{j=1}^{6}
		\eta_j e^{i\mathbf{k}\cdot\boldsymbol{\delta}_j},
		\label{eq:pairing_form_factor}
	\end{equation}
	where $\boldsymbol{\delta}_j$ denotes the six symmetry-related bond vectors within a given neighbor shell, and $\eta_j$ are symmetry-dependent internal bond phases that encode the pairing symmetry as
	$\eta_j^{(l)}=e^{il\theta_j}$~\cite{Maitra2001,Wu2013}. Here, $l$ denotes the relative angular momentum of the pairing symmetry and $\theta_j$ is the polar angle of the $j$th bond with respect to the $x$ axis. For non-onsite pairing, $l=0,\,1,\,2$ and $3$ respectively correspond to extended $s$-wave, $p$-wave, $d$-wave and $f$-wave symmetries. For chiral pairing symmetries, such as $p\pm ip$ corresponds to $l = \pm 1$ and $d\pm id$ corresponds to $l=\pm 2$.\\
	The bond angles for NN and NNN bonds, respectively, are given by
	\begin{equation}
		\theta_j^{\rm NN}=\frac{\pi}{3}(j-1),
		\qquad
		\theta_j^{\rm NNN}=\frac{\pi}{6}(2j-1).
		\label{eq:bond_angles}
	\end{equation}
	The quasiparticle spectrum is obtained by solving
	\begin{equation}
		H_{\mathrm{BdG}}(\mathbf{k})
		\Phi_n(\mathbf{k})
		=
		E_n(\mathbf{k})
		\Phi_n(\mathbf{k}),
		\label{eq:BdG_eigen}
	\end{equation}
	where
	\(
	\Phi_n(\mathbf{k})
	=
	(u_{n\uparrow},
	u_{n\downarrow},
	v_{n\uparrow},
	v_{n\downarrow})^{T}
	\)
	is the Bogoliubov quasiparticle amplitudes, and $E_n$ are the quasiparticle eigen values. 
	Owing to particle-hole symmetry, only the positive-energy quasiparticle solutions are sufficient in the evaluation of the correlation functions in the superconducting state; negative energy solutions are same in magnitude of the positive energy solutions and the corresponding amplitudes can be obtained by using the transformations: $u_{n\sigma} \to v^*_{n\sigma},\, v_{n,\sigma} = -u^*_{n,\sigma}$, where $\sigma=\uparrow,\downarrow$.

	The superconducting correlations are encoded in the anomalous expectation values of the fermionic operators: 
	\begin{equation}
		F_s(\mathbf{k})
		=\dfrac{1}{2}
		\langle
		c_{\mathbf{k}\uparrow}
		c_{-\mathbf{k}\downarrow}
		-
		c_{\mathbf{k}\downarrow}
		c_{-\mathbf{k}\uparrow}
		\rangle,
	\end{equation}
   for the singlet channel. In terms of the BdG eigenvectors, $F_s(\mathbf{k})$ is given by
	\begin{equation}
		\begin{split}
			F_s(\mathbf{k})
			&= \dfrac{1}{2}\sum_{E_n>0}\Bigl[ u_{n\uparrow}(\mathbf{k}) v^{*}_{n\downarrow}(\mathbf{k}) \\
			&\qquad\qquad
			- u_{n\downarrow}(\mathbf{k}) v^{*}_{n\uparrow}(\mathbf{k}) \Bigr] \tanh\left( \frac{E_n(\mathbf{k})}{2k_B T} \right).
			\label{eq:Fs}
		\end{split}
	\end{equation}
	For equal-spin triplet pairing, the anomalous correlators are
	\begin{align}
		F_{\uparrow\uparrow}(\mathbf{k})
		&=
		\langle
		c_{\mathbf{k}\uparrow}
		c_{-\mathbf{k}\uparrow}
		\rangle =
		\sum_{E_n >0}
		u_{n\uparrow}(\mathbf{k})
		v^{*}_{n\uparrow}(\mathbf{k})
		\tanh\left(
		\frac{E_n(\mathbf{k})}{2k_B T}
		\right),\label{eq:Fupup}
		\\
		F_{\downarrow\downarrow}(\mathbf{k})
		&=
		\langle
		c_{\mathbf{k}\downarrow}
		c_{-\mathbf{k}\downarrow}
		\rangle =
		\sum_{E_n>0}
		u_{n\downarrow}(\mathbf{k})
		v^{*}_{n\downarrow}(\mathbf{k})
		\tanh\left(
		\frac{E_n(\mathbf{k})}{2k_B T}
		\right).\label{eq:Fdowndown}
	\end{align}
	
	The superconducting order parameters are obtained self-consistently by restricting the momentum summation to electronic states within an energy window set by the Debye energy $\omega_D$ around the Fermi level. Therefore, the singlet and equal-spin triplet components are then given by,
	\begin{align}
		\Delta_s &=
		\frac{V_s}{N_D}\sum_{\mathbf{k}\in \mathbf{k}_D}
		\phi_s^*(\mathbf{k})F_s(\mathbf{k}), \label{eq:singlet_gap}\\[-2pt]
		\Delta_{\sigma\sigma} &=
		\frac{V_t}{N_D}\sum_{\mathbf{k}\in \mathbf{k}_D}
		\phi_t^*(\mathbf{k})F_{\sigma\sigma}(\mathbf{k}). \label{eq:triplet_gap}
	\end{align}
	 Here, $V_s$ and $V_t$ denote the strengths of singlet and triplet interaction channels respectively, $\mathbf{k}_D$ represents the cutoff momentum such that $|E_n(\mathbf{k}_D)|\leq \omega_D$, and $N_D$ is the number of $\mathbf{k}$ points within the Debye shell.
	
	To quantify the momentum-space origin of the superconducting order parameter, we further evaluate the partial contributions from regions surrounding the high-symmetry points $\Gamma$, $K$, and $K'$,
	\begin{align}
		\Delta_s^{(P)}
		&= \frac{V_s}{N_P} \sum_{\mathbf{k}\in \mathbf{k}_P} \phi_s^{*}(\mathbf{k}) F_s(\mathbf{k}),
		\label{eq:patch_singlet}\\
		\Delta_{\sigma\sigma}^{(P)}
		&= \frac{V_t}{N_P} \sum_{\mathbf{k}\in \mathbf{k}_P} \phi_t^{*}(\mathbf{k}) F_{\sigma\sigma}(\mathbf{k}).
		\label{eq:patch_triplet}
	\end{align}
	where $P=\Gamma,K,K'$ denotes the corresponding momentum-space patch and $N_P$ is the number of $\mathbf{k}$ points within each patch. By construction, the patches are chosen to partition the Debye shell without overlap. The full order parameters in Eqs.\eqref{eq:singlet_gap} and \eqref{eq:patch_triplet} can also be recovered
	using the weighted sum
	\begin{equation}
		\Delta_s
		=
		\sum_P
		\frac{N_P}{N_D}
		\Delta_s^{(P)},
		\qquad
		\Delta_{\sigma\sigma}
		=
		\sum_P
		\frac{N_P}{N_D}
		\Delta_{\sigma\sigma}^{(P)}.
		\label{eq:full_gap_from_patches}
	\end{equation}

	\subsubsection{\textbf{Spin-singlet pairing channels}}
	
	\begin{figure}[b]
		\centering
		
		\begin{tikzpicture}[
			scale=0.48,
			bond/.style={line width=0.8pt, draw=black!75!black},
			site/.style={circle, fill=red!75!black, inner sep=1.8pt},
			center/.style={circle, fill=red!75!black, inner sep=2.0pt},
			phase/.style={font=\scriptsize},
			]
			
			\begin{scope}[xshift=0cm]
				\def\r{1.0}
				
				\foreach \ang in {0,60,120,180,240,300}{
					\draw[bond] (0,0)--(\ang:\r);
					\node[site] at (\ang:\r) {};
				}
				
				\node[center] at (0,0) {};
				
				\foreach \ang in {0,60,120,180,240,300}{
					\node[phase] at (\ang:1.45) {$1$};
				}
				
				\node at (0,-3.0)
				{\scriptsize \textbf{(a)} $\phi^{NN}_{\mathrm{ext}\text{-}s}$};
			\end{scope}

			\begin{scope}[xshift=4.0cm]
				\def\r{1.0}
				
				\foreach \ang in {0,60,120,180,240,300}{
					\draw[bond] (0,0)--(\ang:\r);
					\node[site] at (\ang:\r) {};
				}
				
				\node[center] at (0,0) {};
				
				\node[phase] at (0:1.5) {$1$};
				\node[phase] at (55:1.8)
				{$e^{\pm i\frac{2\pi}{3}}$};
				\node[phase] at (125:1.8)
				{$e^{\mp i\frac{2\pi}{3}}$};
				\node[phase] at (180:1.5) {$1$};
				\node[phase] at (235:1.8)
				{$e^{\pm i\frac{2\pi}{3}}$};
				\node[phase] at (305:1.8)
				{$e^{\mp i\frac{2\pi}{3}}$};
				
				\node at (0,-3.0)
				{\scriptsize \textbf{(b)}
					$\phi_{d}\pm i\phi_{d}$};
			\end{scope}

			\begin{scope}[xshift=8.0cm]
				\def\r{1.7}
				
				\foreach \ang in {30,90,150,210,270,330}{
					\draw[bond] (0,0)--(\ang:\r);
					\node[site] at (\ang:\r) {};
				}
				
				\foreach \ang in {0,60,120,180,240,300}{
					\node[site] at (\ang:1.0) {};
				}
				
				\node[center] at (0,0) {};
				
				\foreach \ang in {30,90,150,210,270,330}{
					\node[phase] at (\ang:2.15) {$1$};
				}
				
				\node at (0,-3.0)
				{\scriptsize \textbf{(c)} $\phi^{NNN}_{\mathrm{ext}\text{-}s}$};
			\end{scope}

			\begin{scope}[xshift=13.0cm]
				\def\r{1.7}
				
				\foreach \ang in {30,90,150,210,270,330}{
					\draw[bond] (0,0)--(\ang:\r);
					\node[site] at (\ang:\r) {};
				}
				
				\foreach \ang in {0,60,120,180,240,300}{
					\node[site] at (\ang:1.0) {};
				}
				
				\node[center] at (0,0) {};
				
				\node[phase] at (35:2.4)
				{$e^{\pm i\frac{\pi}{3}}$};
				\node[phase] at (90:2.2) {$-1$};
				\node[phase] at (145:2.4)
				{$e^{\mp i\frac{\pi}{3}}$};
				\node[phase] at (215:2.4)
				{$e^{\pm i\frac{\pi}{3}}$};
				\node[phase] at (270:2.2) {$-1$};
				\node[phase] at (325:2.4)
				{$e^{\mp i\frac{\pi}{3}}$};
				
				\node at (0,-3.0)
				{\scriptsize \textbf{(d)}
					$\phi_{d'}\pm i\phi_{d'}$};
			\end{scope}
			
		\end{tikzpicture}
		
		\caption{Real-space internal bond phases $\eta_j^{(l)}$ for the four
			spin-singlet pairing channels in monolayer NbSe$_2$: NN extended-$s$
			and chiral $d_{x^2-y^2}\pm i d_{xy}$ pairing [(a) and (b)], and NNN
			extended-$s$ and chiral $d'_{x^2-y^2}\pm i d'_{xy}$ pairing [(c) and
			(d)]. The red circles and bonds denote the Nb sites and corresponding
			pairing bonds, respectively.}
		
		\label{fig:singlet_bond_pairing}
	\end{figure}
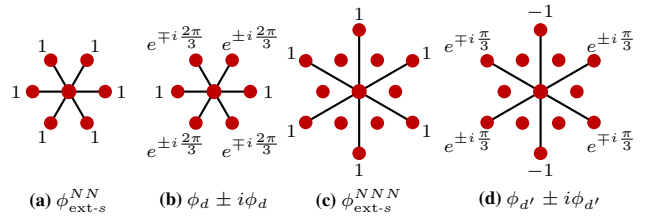
	
	\begin{table}[t]
		\caption{Momentum-space basis functions for the bond-dependent spin-singlet pairing channels in monolayer NbSe$_2$, where $\alpha=k_x/2$ and $\beta=\sqrt{3}k_y/2$.}
		\label{tab:singlet_basis}
		\centering
		\begin{tabular}{lcc}
			\hline\hline
			\noalign{\vskip 1mm}
			Pairing channel & NN basis function & NNN basis function \\
			\noalign{\vskip 1mm}
			\hline
			\noalign{\vskip 1mm}
			Extended-$s$ &
			$\cos(2\alpha)+2\cos\alpha\cos\beta$ &
			$\cos(2\beta) +2\cos(3\alpha)\cos\beta$ \\[2mm]
			
			$d_{x^2-y^2}$ &
			$\cos\alpha\cos\beta - \cos(2\alpha)$ &
			$\cos(2\beta) - \cos(3\alpha)\cos\beta$ \\[2mm]
			
			$d_{xy}$ &
			$\sqrt{3}\sin\alpha\sin\beta$ &
			$\sqrt{3}\sin(3\alpha)\sin\beta$ \\[2mm]
			
			Chiral--$d$ &
			$\phi_{d_{x^2-y^2}}\pm i\phi_{d_{xy}}$ &
			$\phi_{d'_{x^2-y^2}}\pm i\phi_{d'_{xy}}$ \\
			\hline\hline
		\end{tabular}
	\end{table}

	The momentum-space basis functions for both NN and NNN pairings are tabulated in Table~\ref{tab:singlet_basis}. The complex combination of the two basis functions gives rise to the
	chiral $d\pm id$ state, which spontaneously breaks time-reversal symmetry. The spin-singlet $s$-wave channel with onsite pairing is described by the trivial isotropic form factor $\phi_s(\mathbf{k})=1$. The remaining bond-dependent single pairing channels are classified according to the irreducible
	representations of the crystal point group $D_{3h}$. The extended-$s$
	channel belongs to the totally symmetric representation $A_1'$, whereas
	$\left(d_{x^2-y^2},\,d_{xy}\right)$ span the two-dimensional representation
	$E'$. The corresponding real-space bond phases are illustrated in
	Fig.~\ref{fig:singlet_bond_pairing}.

	\subsubsection{\textbf{Spin-triplet pairing channels}}
	
	The equal-spin triplet components are described by
	$\Delta_{\uparrow\uparrow}\phi_t(\mathbf{k})$ and
	$\Delta_{\downarrow\downarrow}\phi_t(\mathbf{k})$, where
	$\phi_t(\mathbf{k})$ is an odd-parity pairing basis function on the
	triangular lattice. Although monolayer NbSe$_2$ is predominantly a
	spin-singlet superconductor, intrinsic Ising spin--orbit coupling
	together with an in-plane Zeeman field induces equal-spin triplet
	correlations.
	
	\begin{figure}[hbt]
		\centering
		
		\begin{tikzpicture}[
			scale=0.48,
			bond/.style={line width=0.8pt, draw=black!75!black},
			site/.style={circle, fill=red!75!black, inner sep=1.8pt},
			center/.style={circle, fill=red!75!black, inner sep=2.0pt},
			phase/.style={font=\scriptsize},
			]
			
			\begin{scope}[xshift=0cm]
				\def\r{1.0}
				
				\foreach \ang in {0,60,120,180,240,300}{
					\draw[bond] (0,0)--(\ang:\r);
					\node[site] at (\ang:\r) {};
				}
				
				\node[center] at (0,0) {};
				
				\node[phase] at (0:1.55) {$+1$};
				\node[phase] at (50:1.85)
				{$e^{\pm i\frac{\pi}{3}}$};
				\node[phase] at (120:1.7)
				{$-e^{\mp i\frac{\pi}{3}}$};
				\node[phase] at (180:1.55) {$-1$};
				\node[phase] at (240:1.8)
				{$-e^{\pm i\frac{\pi}{3}}$};
				\node[phase] at (310:1.85)
				{$e^{\mp i\frac{\pi}{3}}$};
				
				\node at (0,-3.0)
				{\scriptsize \textbf{(a)}
					$\phi_{p}\pm i\phi_{p}$};
			\end{scope}

			\begin{scope}[xshift=4.0cm]
				\def\r{1.0}
				
				\foreach \ang in {0,60,120,180,240,300}{
					\draw[bond] (0,0)--(\ang:\r);
					\node[site] at (\ang:\r) {};
				}
				
				\node[center] at (0,0) {};
				
				\node[phase] at (0:1.55) {$+1$};
				\node[phase] at (65:1.55) {$-1$};
				\node[phase] at (120:1.6) {$+1$};
				\node[phase] at (180:1.55) {$-1$};
				\node[phase] at (240:1.7) {$+1$};
				\node[phase] at (290:1.6) {$-1$};
				
				\node at (0,-3.0)
				{\scriptsize \textbf{(b)}
					$\phi_{f}$};
			\end{scope}

			\begin{scope}[xshift=8.0cm]
				\def\r{1.7}
				
				\foreach \ang in {30,90,150,210,270,330}{
					\draw[bond] (0,0)--(\ang:\r);
					\node[site] at (\ang:\r) {};
				}
				
				\foreach \ang in {0,60,120,180,240,300}{
					\node[site] at (\ang:1.0) {};
				}
				
				\node[center] at (0,0) {};
				
				\node[phase] at (35:2.4)
				{$e^{\pm i\frac{\pi}{6}}$};
				\node[phase] at (90:2.2) {$\pm i$};
				\node[phase] at (140:2.25)
				{$-e^{\mp i\frac{\pi}{6}}$};
				\node[phase] at (220:2.25)
				{$-e^{\pm i\frac{\pi}{6}}$};
				\node[phase] at (270:2.2) {$\mp i$};
				\node[phase] at (325:2.4)
				{$e^{\mp i\frac{\pi}{6}}$};
				
				\node at (0,-3.0)
				{\scriptsize \textbf{(c)}
					$\phi_{p^\prime}\pm i\phi_{p^\prime}$};
			\end{scope}

			\begin{scope}[xshift=13.0cm]
				\def\r{1.7}
				
				\foreach \ang in {30,90,150,210,270,330}{
					\draw[bond] (0,0)--(\ang:\r);
					\node[site] at (\ang:\r) {};
				}
				
				\foreach \ang in {0,60,120,180,240,300}{
					\node[site] at (\ang:1.0) {};
				}
				
				\node[center] at (0,0) {};
				
				\node[phase] at (35:2.2) {$+i$};
				\node[phase] at (90:2.2) {$-i$};
				\node[phase] at (145:2.2) {$+i$};
				\node[phase] at (215:2.2) {$-i$};
				\node[phase] at (270:2.2) {$+i$};
				\node[phase] at (325:2.2) {$-i$};
				
				\node at (0,-3.0)
				{\scriptsize \textbf{(d)}
					$\phi_{f^\prime}$};
			\end{scope}
			
		\end{tikzpicture}
		
		\caption{Real-space internal bond phases $\eta_j^{(l)}$ for the four
			spin-triplet pairing channels in monolayer NbSe$_2$: NN chiral
			$p_x\pm ip_y$ and $f_{x(x^2-3y^2)}$ pairing [(a) and (b)], and NNN
			chiral $p^\prime_x\pm ip^\prime_y$ and
			$f^\prime_{y(3x^2-y^2)}$ pairing [(c) and (d)]. The red circles
			and bonds denote the Nb sites and corresponding pairing bonds,
			respectively.}
		
		\label{fig:triplet_bond_pairing}
	\end{figure}
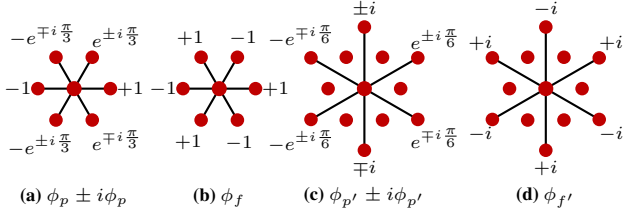
	
	\begin{table}[hbt]
		\caption{Momentum-space basis functions for the bond-dependent equal-spin triplet pairing channels in monolayer NbSe$_2$, where $\alpha=k_x/2$ and $\beta=\sqrt{3}k_y/2$.}
		\label{tab:triplet_basis}
		\centering
		\begin{tabular}{lcc}
			\hline\hline
			\noalign{\vskip 1mm}
			Pairing channel & NN basis function & NNN basis function \\
			\noalign{\vskip 1mm}
			\hline
			\noalign{\vskip 1mm}
			$p_x$  &
			$\sin(2\alpha)+\sin\alpha\cos\beta$ &
			$\sqrt{3}\sin(3\alpha)\cos\beta$ \\[2mm]
			
			$p_y$  &
			$\sqrt{3}\cos\alpha\,\sin\beta$ &
			$\cos(3\alpha)\sin\beta+\sin(2\beta)$ \\[2mm]
			
			Chiral--$p$ &
			$\phi_{p_x}\pm i\phi_{p_y}$ &
			$\phi_{p'_x}\pm i\phi_{p'_y}$ \\[2mm]
			
			$f_{x(x^2-3y^2)}$  &
			$2\sin\alpha\cos\beta - \sin(2\alpha)$ &
			--- \\[2mm]
			
			$f^\prime_{y(3x^2-y^2)}$ &
			--- &
			$\sin(2\beta)-2\cos(3\alpha)\sin\beta$ \\[1mm]
			
			\hline\hline
		\end{tabular}
	\end{table}
	
	The triplet basis functions are classified according to the irreducible
	representations of the crystal point group $D_{3h}$. 
	The basis functions for $p_x$ and $p_y$ channels belong to the $E'$ representation, and their complex
	combinations give the chiral $p_x\pm ip_y$ state. 
	The basis functions for $f_{x(x^2-3y^2)}$ and $f^\prime_{y(3x^2-y^2)}$ channels belong to the $A_1'$ and $A_2'$
	representations, respectively. Their real-space bond phases are shown
	in Fig.~\ref{fig:triplet_bond_pairing}, while the corresponding
	momentum-space basis functions are summarized in
	Table~\ref{tab:triplet_basis}.
	
	\section{\label{sec:IV} Self-Consistent \NoCaseChange{BdG} Solutions}
	
	We perform all self-consistent calculations using a $2000\times2000$ Monkhorst-Pack momentum grid at a temperature of $T=1.5$ K with a Debye cutoff energy $\hbar\omega_D=50$ meV. The chemical potential is fixed at the physical value, $\mu=-31.4$ meV, corresponding to the tight-binding model of monolayer NbSe$_2$ \cite{Bobkov_2024_A}. For each pairing channel, the singlet pairing interaction is chosen such that the self-consistent singlet gap satisfies $|\Delta_s|\approx0.8$ meV, consistent with the experimentally reported superconducting gap of $0.5$--$1.0$ meV \cite{Kuzmanovic2022}. This corresponds to $V_s=33.2$ meV for onsite pairing, $V_s=22.8$ meV for NN pairing, and $V_s=17.1$ meV for NNN pairing. The triplet pairing interaction is set to zero ($V_t=0$) for calculations of estimating $T_c$ and obtaining the phase diagrams in ($h_x,\mu$) plane. Triplet channel is further opened $(V_t \neq 0)$ to obtain the induced triplet order parameter in presence of an in-plane magnetic field.
	
	The BdG Hamiltonian is diagonalized at each momentum according to Eq.~\eqref{eq:BdG_eigen}, and the resulting quasiparticle eigenvalues and eigenvectors are used to evaluate the anomalous correlators in Eqs.~\eqref{eq:Fs}--\eqref{eq:Fdowndown}. The singlet and equal-spin triplet order parameters are then updated using the self-consistency equations, Eqs.~\eqref{eq:singlet_gap} and \eqref{eq:triplet_gap}, respectively, and iterated until convergence. For finite $V_t$, the singlet and triplet components are solved simultaneously.
	
	To resolve the momentum-space contributions to the superconducting order, we evaluate the self-consistency equations separately within patches around the high-symmetry points $\Gamma$, $K$, and $K'$, using Eqs.~\eqref{eq:patch_singlet} and \eqref{eq:patch_triplet}. For field-induced equal-spin triplet pairing, we define the total triplet order parameter as $ \Delta_t=\Delta_{\uparrow\uparrow} = \Delta_{\downarrow\downarrow}.$
	
	\subsection{Onsite $s$-wave pairing}
	
	\begin{figure}[t]
		\centering
		\begin{minipage}[t]{0.49\linewidth}
			\centering
			\includegraphics[width=\linewidth]{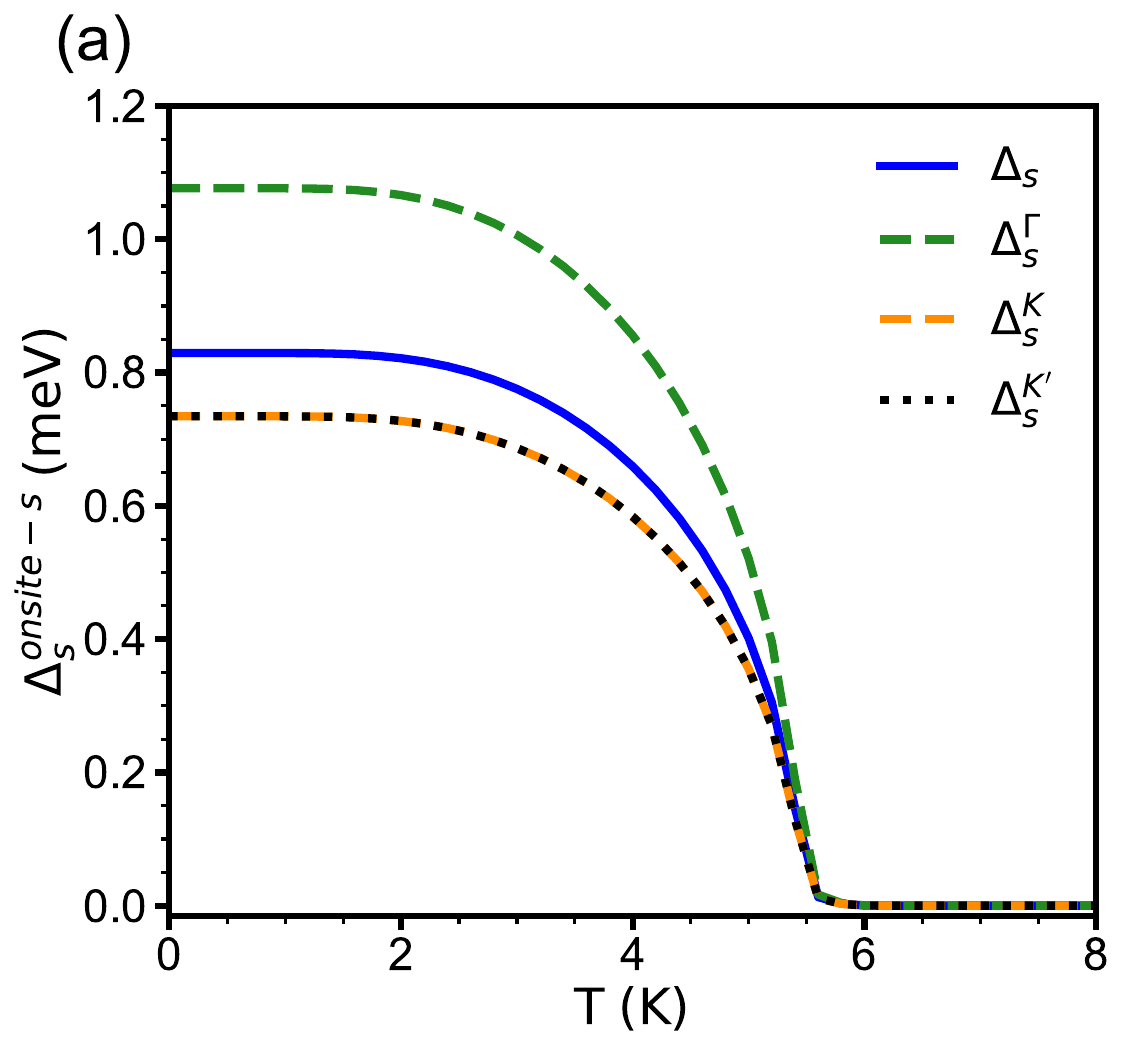}
		\end{minipage}\hfill
		\begin{minipage}[t]{0.49\linewidth}
			\centering
			\includegraphics[width=\linewidth]{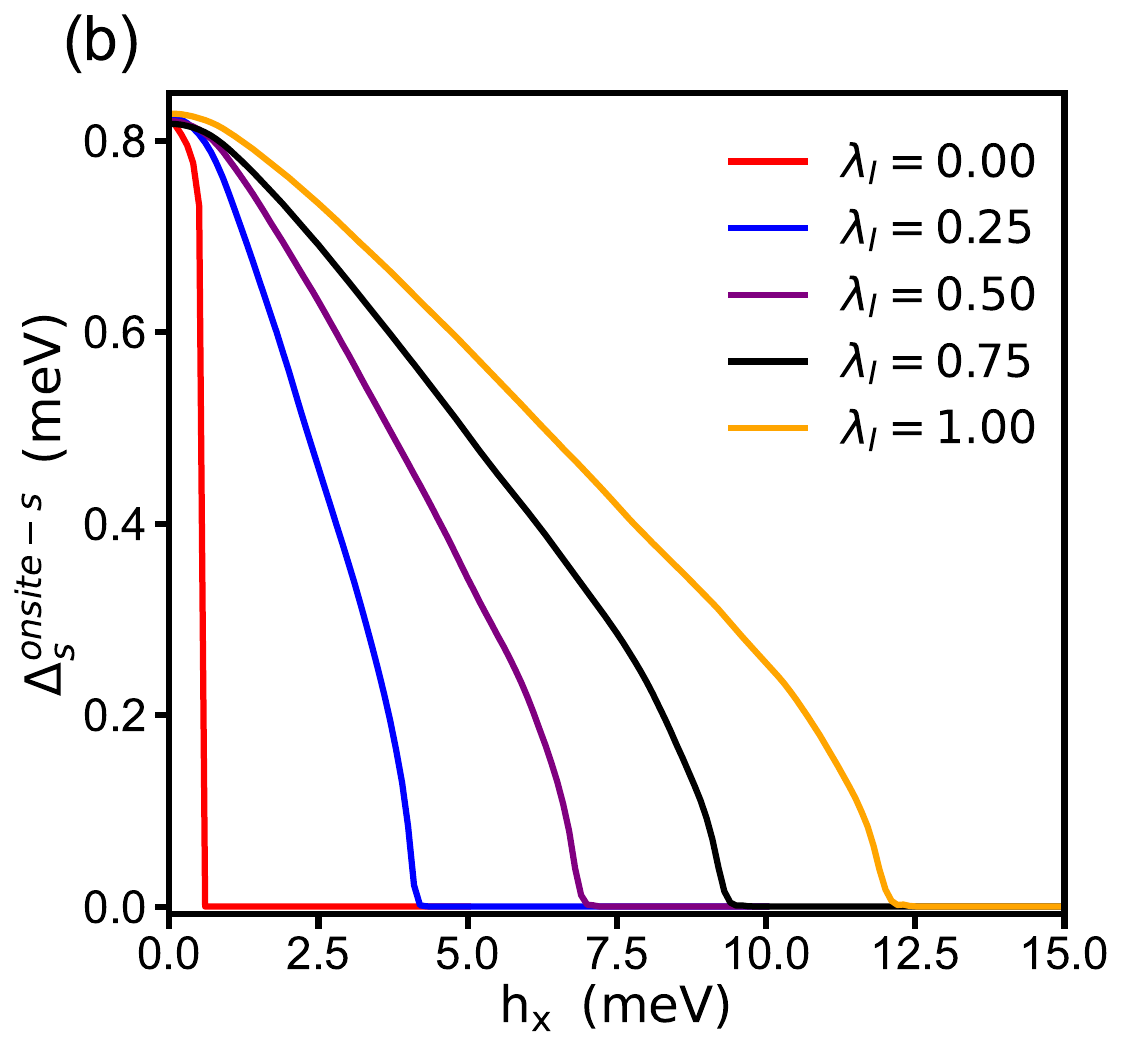}
		\end{minipage}
		\caption{(a) Temperature dependence of the self-consistent onsite $s$-wave gap around the $\Gamma$ and $K/K'$ regions of the Brillouin zone. (b) Onsite $s$-wave order parameter as a function of the in-plane Zeeman field for different Ising SOC strengths.}
		\label{fig:onsite_s}
	\end{figure}
	
	We first consider the conventional onsite spin-singlet superconducting state in monolayer NbSe$_2$. 
	As shown in Fig.~\ref{fig:onsite_s}(a), the superconducting gap decreases monotonically with increasing temperature and vanishes at $T_c\approx5.8$ K. Throughout the superconducting phase, the regions near $\Gamma$, $K$, and $K'$ points contribute appreciably to the superconducting order parameter, with the region near $\Gamma$ providing the dominant contribution over the entire temperature range. Since the onsite $s$-wave basis function is momentum independent, $\phi_s(\mathbf{k})=1$, the relative contributions from the regions near $\Gamma$, $K$, and $K'$ points are governed primarily by the distribution of electronic states within the Debye energy window.
	
	The magnetic-field dependence of the order parameter is shown in Fig.~\ref{fig:onsite_s}(b). In the absence of Ising SOC ($\lambda_I=0$), superconductivity is rapidly suppressed by conventional Pauli pair breaking mechanism. Increasing the Ising SOC progressively weakens the pair-breaking effect, shifting the superconducting transition to higher magnetic fields. We next examine the field-induced equal-spin triplet component.

	\subsubsection{\textbf{Field-induced equal-spin $f$-wave triplet pairing}}
	
	The interplay between Ising spin-orbit coupling and an in-plane Zeeman field induces an odd-parity equal-spin triplet component in addition to the onsite spin-singlet state. To determine the symmetry of the induced pairing, we simultaneously solve the singlet and triplet gap equations self-consistently for all symmetry-allowed NN and NNN equal-spin triplet channels. We find that the onsite $s$-wave state selectively induces a nearest-neighbor $f$-wave triplet component, while the NN chiral $p\pm ip$ channels, and the NNN $f^\prime$-wave and chiral $p^\prime\pm ip^\prime$ channels remain absent throughout the superconducting phase. The corresponding results for the other triplet channels are summarized in Appendix~\ref{app:onsite_s_triplet_channels}.
	
	Figure~\ref{fig:onsite_fwave}(a) shows the evolution of the singlet and induced equal-spin $f$-wave order parameters with in-plane magnetic field. The triplet component is absent at zero field and emerges through the combined action of the Zeeman field and Ising spin-orbit coupling. It increases with field, reaches a maximum at an intermediate field, and vanishes together with the singlet order parameter at the critical field, confirming that the triplet state is induced by the parent singlet superconductivity.
	
	\begin{figure}[t]
		\centering
		
		\begin{minipage}[t]{0.49\linewidth}
			\centering
			\includegraphics[width=\linewidth]{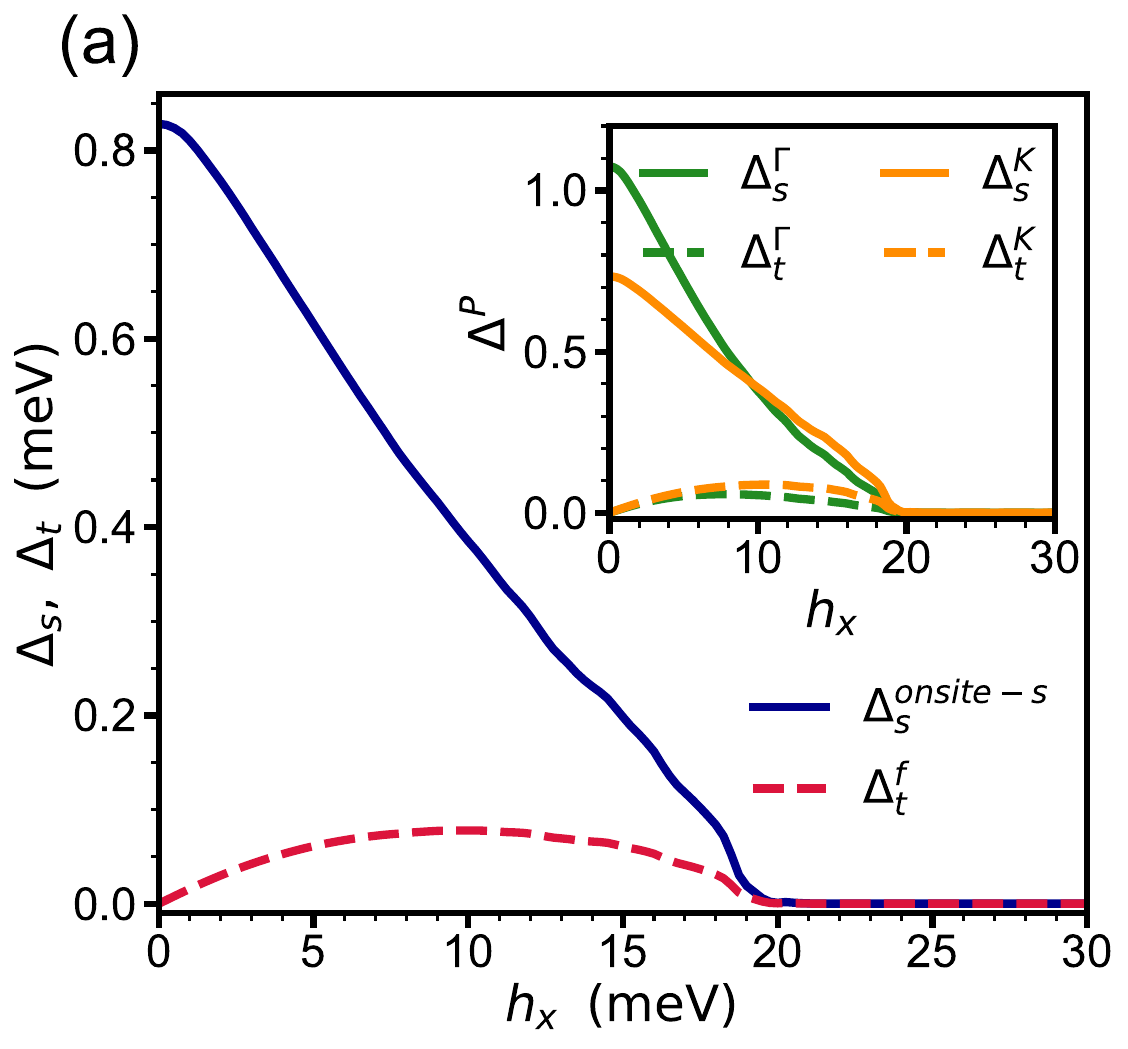}
		\end{minipage}
		\hfill
		\begin{minipage}[t]{0.49\linewidth}
			\centering
			\includegraphics[width=\linewidth]{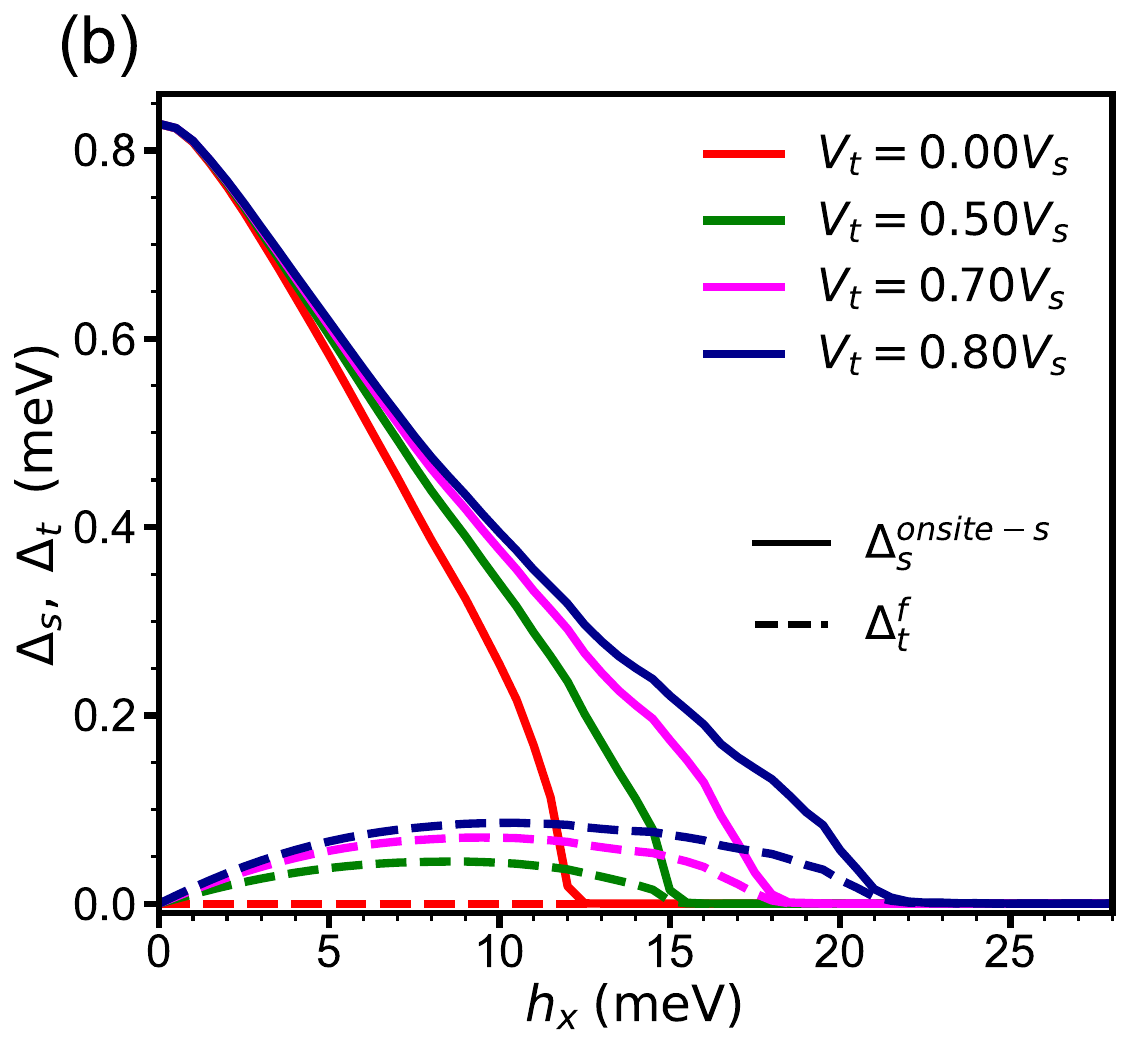}
		\end{minipage}
		
		\caption{(a) Self-consistent onsite-$s$ singlet and induced equal-spin NN $f$-wave triplet order parameters as functions of the in-plane Zeeman field for $V_t=0.75V_s$. The inset compares the $\Gamma$- and $K$-resolved order parameters. (b) Field dependence of the singlet and induced triplet order parameters for different interaction strength $V_t$ in triplet pairing channels.}
		\label{fig:onsite_fwave}
	\end{figure}
	
	The inset of Fig.~\ref{fig:onsite_fwave}(a) compares the momentum-resolved singlet and triplet order parameters from the near $\Gamma$ and $K$ points. At low fields, the singlet contribution is larger around the regions near $\Gamma$ pocket, whereas the induced $f$-wave triplet component receives comparable contributions from both regions, with a slightly larger contribution from the $K$ valleys over the relevant field range. Thus, the momentum-space distribution of the induced triplet component differs from that of the parent onsite singlet state, showing a modest enhancement near the $K$ valleys, where the Ising spin splitting is pronounced.
	
	\begin{figure}[t!]
		\centering
		
		\begin{minipage}[t]{0.24\linewidth}
			\centering
			\includegraphics[height=3.4cm]{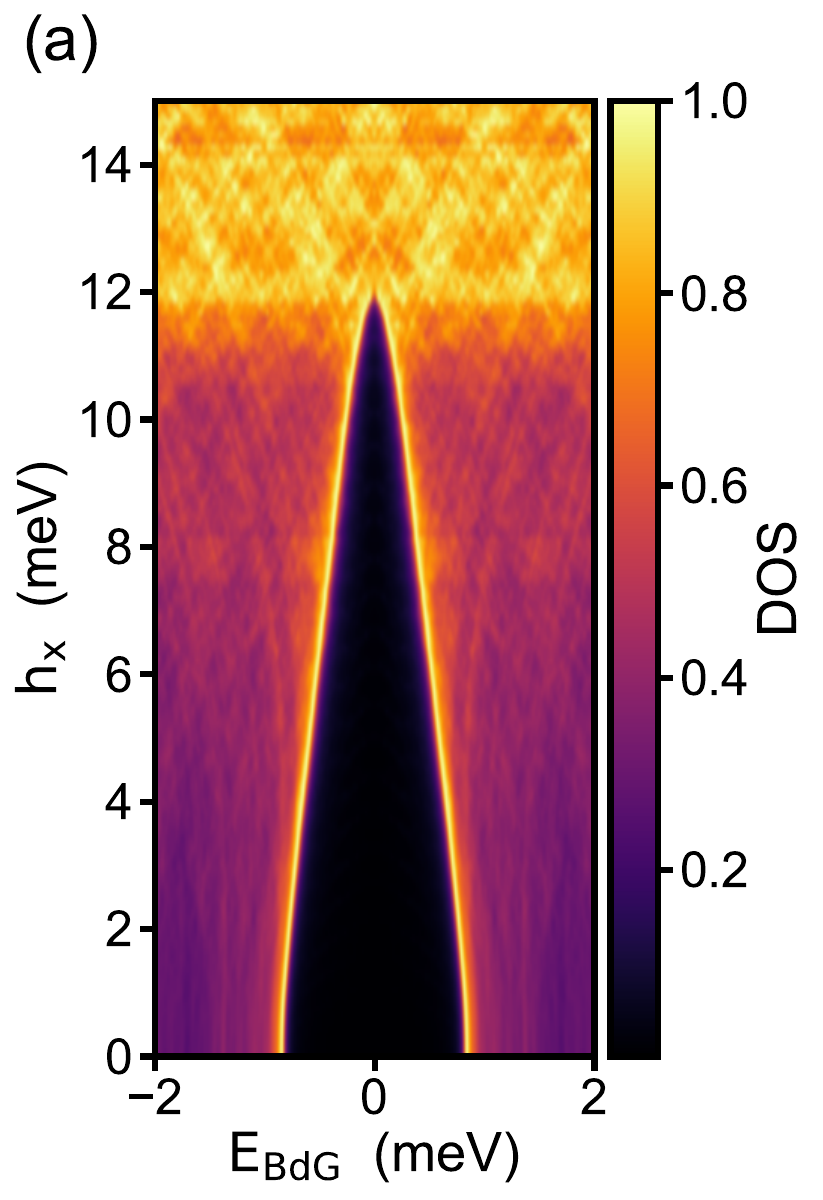}
		\end{minipage}
		\hfill
		\begin{minipage}[t]{0.24\linewidth}
			\centering
			\includegraphics[height=3.4cm]{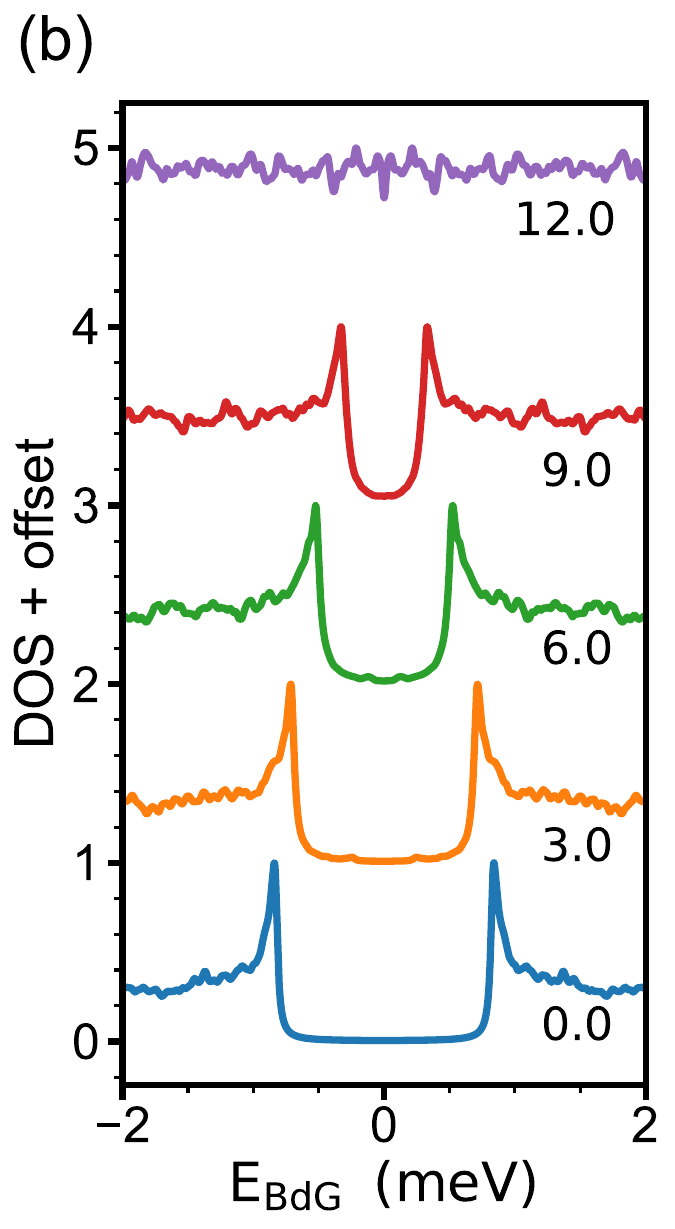}
		\end{minipage}
		\hfill
		\begin{minipage}[t]{0.24\linewidth}
			\centering
			\includegraphics[height=3.4cm]{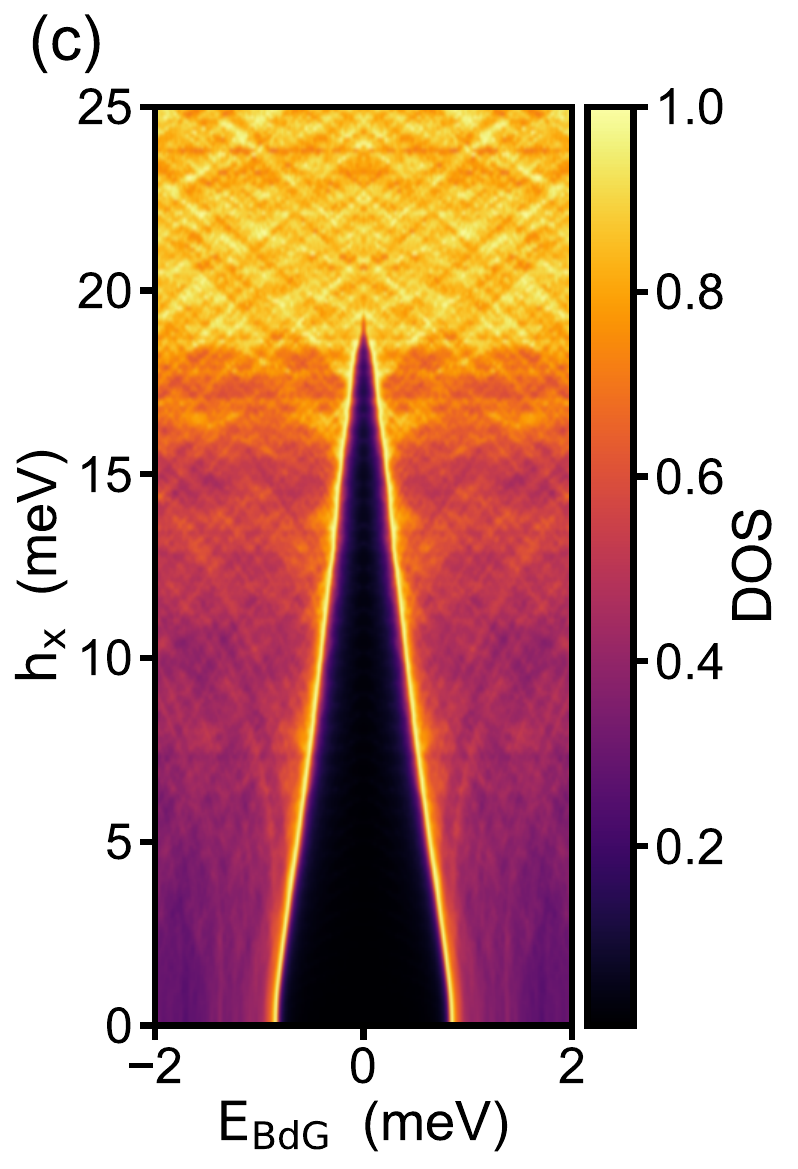}
		\end{minipage}
		\hfill
		\begin{minipage}[t]{0.24\linewidth}
			\centering
			\includegraphics[height=3.4cm]{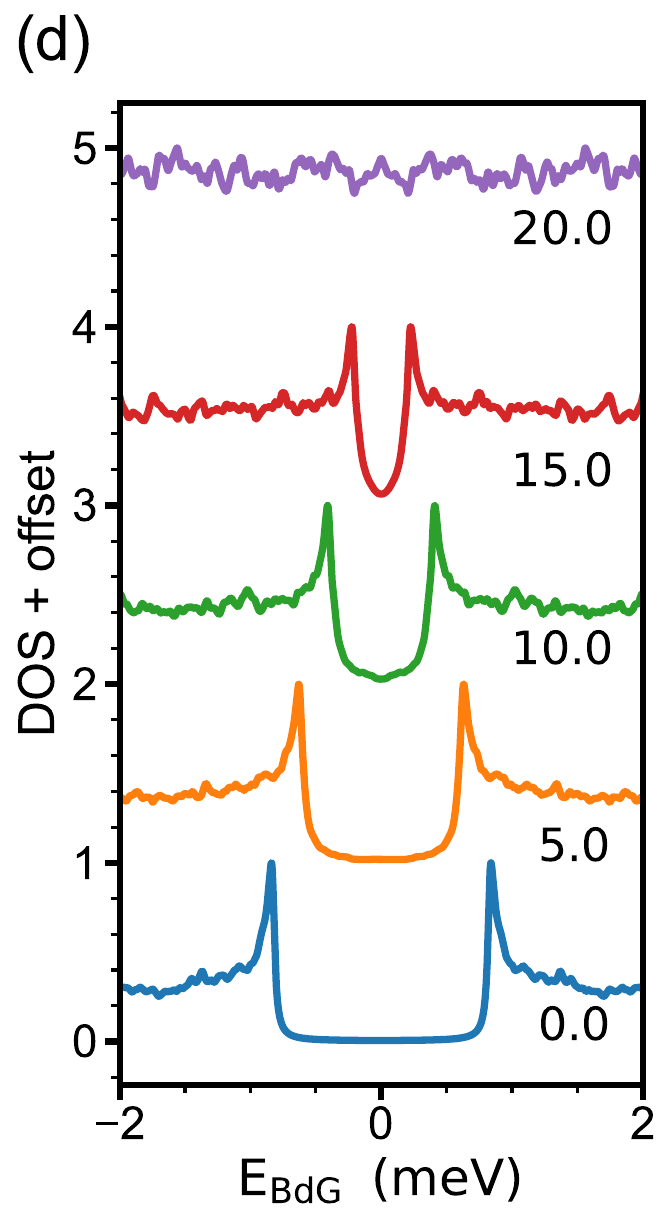}
		\end{minipage}
		
	\caption{	Quasiparticle density of states (DOS) of the onsite $s$-wave superconducting state under an in-plane Zeeman field. (a),(c) Field-dependent DOS heatmaps for $V_t=0$ and $V_t=0.75V_s$, respectively. (b),(d) DOS at selected values of $h_x$ (in meV) corresponding to (a) and (c), respectively; successive curves are vertically offset for clarity.}
		\label{fig:onsite_dos}
	\end{figure}
	
	Figure~\ref{fig:onsite_fwave}(b) shows that increasing the intrinsic triplet interaction enhances the induced $f$-wave order parameter and shifts the superconducting critical field to higher values. This demonstrates that an attractive triplet interaction cooperates with the field-induced singlet-triplet conversion to stabilize the mixed-parity superconducting state.
	
	\begin{figure*}[t]
		\centering
		
		\begin{minipage}[t]{0.32\textwidth}
			\centering
			\includegraphics[width=\linewidth]{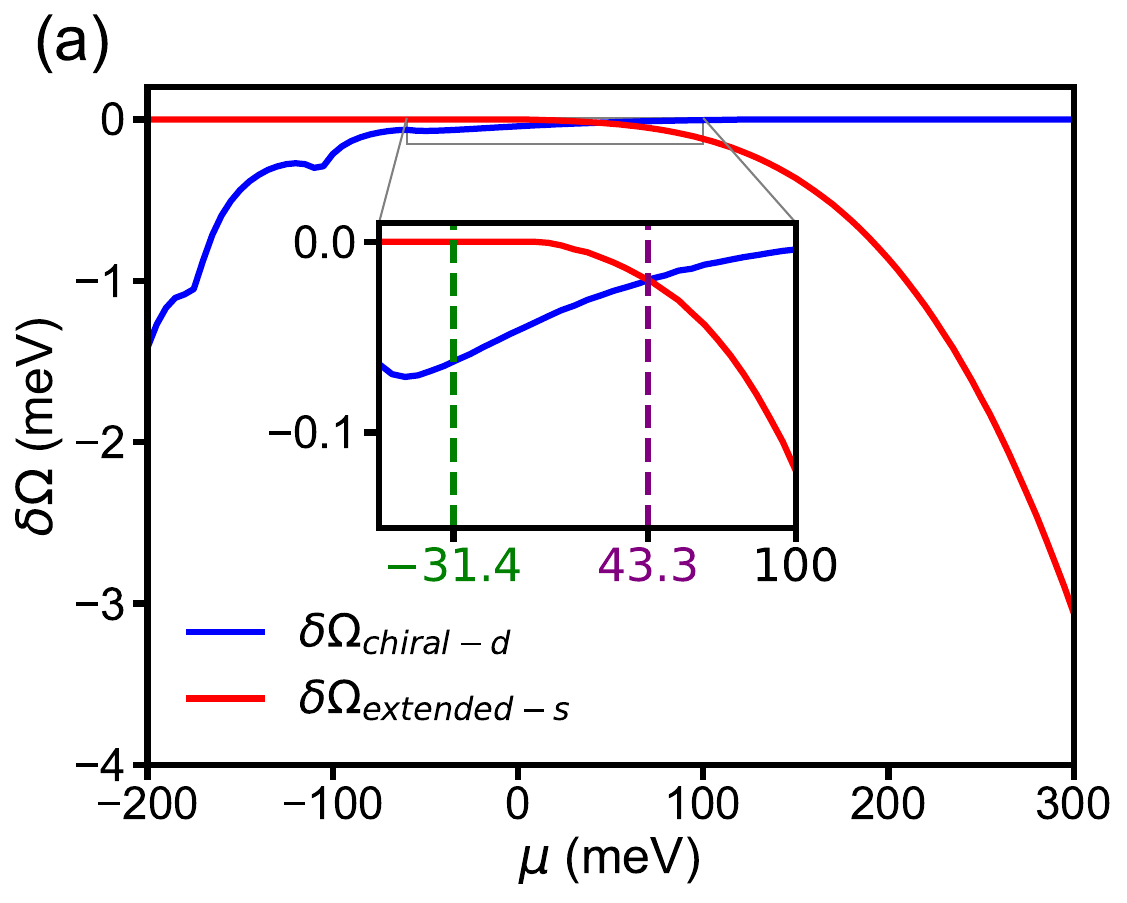}
		\end{minipage}
		\hfill
		\begin{minipage}[t]{0.32\textwidth}
			\centering
			\includegraphics[width=\linewidth]{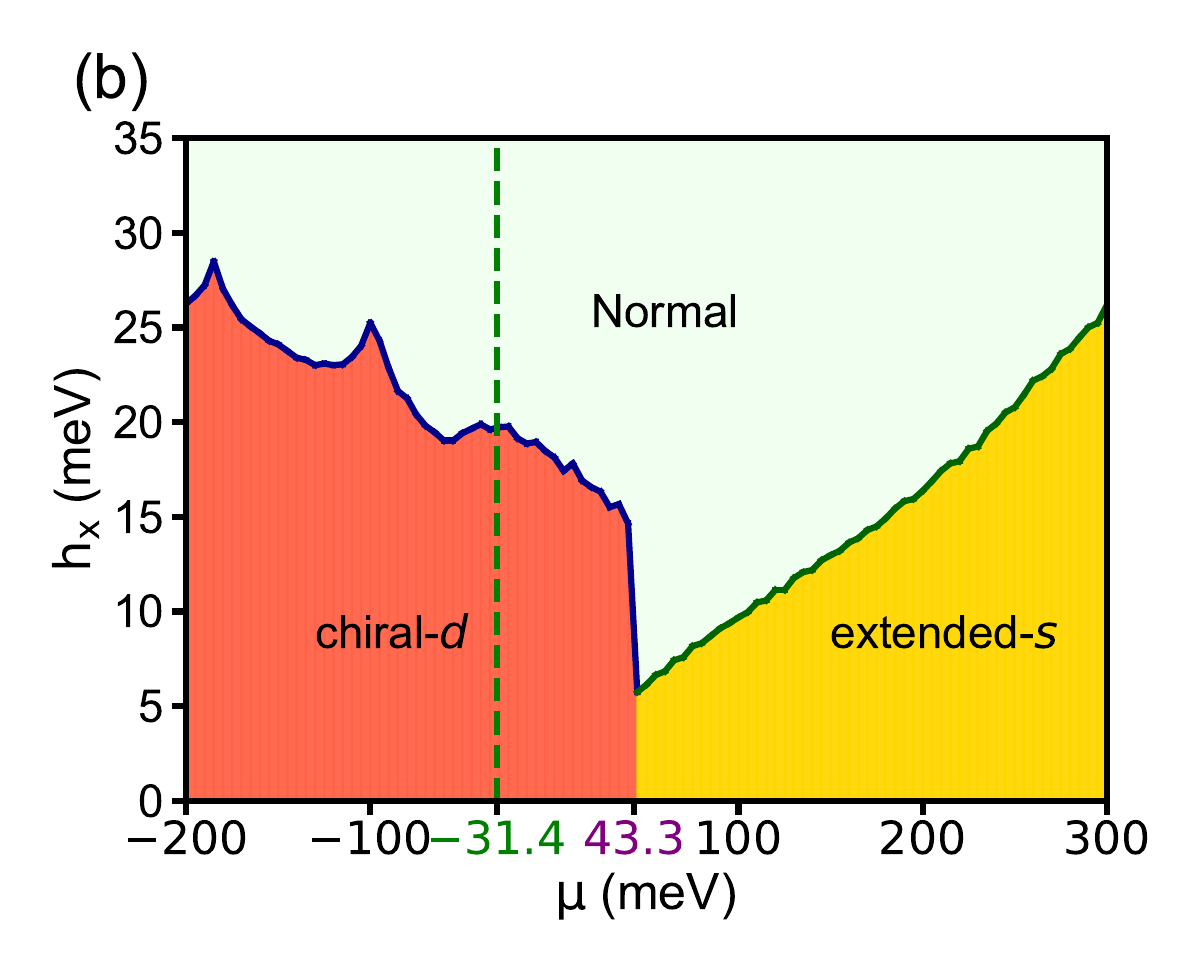}
		\end{minipage}
		\hfill
		\begin{minipage}[t]{0.32\textwidth}
			\centering
			\includegraphics[width=\linewidth]{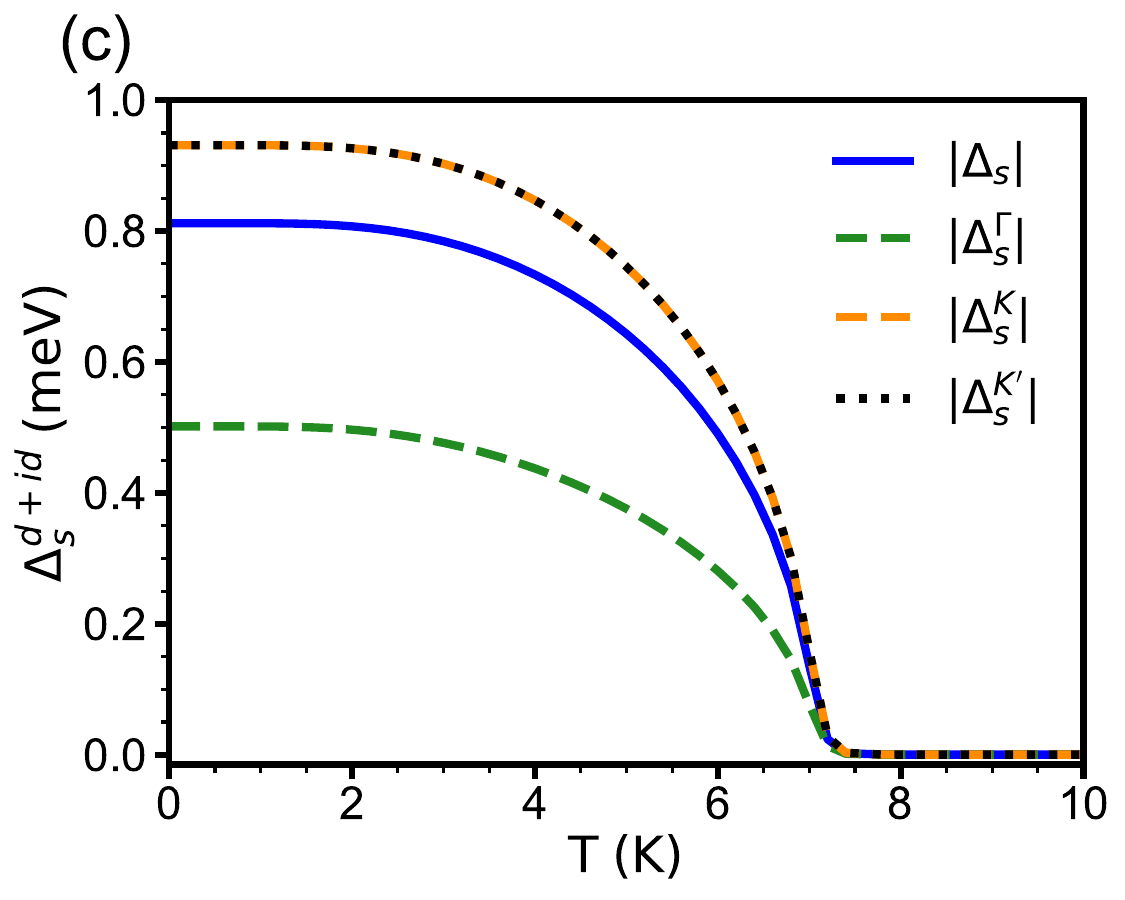}
		\end{minipage}
		
		\caption{
			(a) Condensation energies, $\delta\Omega$, of the chiral-$d$ and extended-$s$ states as functions of chemical potential, with the inset showing their crossover region.
			(b) Superconducting phase diagram in the $(h_x,\mu)$ plane, showing the regions of chiral-$d$, extended-$s$, and normal states. The vertical dashed line indicates the  
			chemical potential of monolayer NbSe$_2$.
			(c) Temperature evolution of the self-consistent $d+id$ order parameter, together with its contributions from the $\Gamma$  and $K/K'$ valleys.
		}
		\label{fig:NN_pairing}
	\end{figure*}
	
	The evolution of the quasiparticle density of states with $h_x$ is shown in Fig.~\ref{fig:onsite_dos}. For the pure onsite singlet state ($V_t=0$), the superconducting gap gradually decreases with increasing in-plane field and closes at the critical field, accompanied by the disappearance of the coherence peaks [Figs.~\ref{fig:onsite_dos}(a) and \ref{fig:onsite_dos}(b)]. When a finite triplet interaction ($V_t=0.75V_s$) is included, the overall evolution of the spectrum remains qualitatively unchanged, but the superconducting gap persists to substantially higher magnetic fields [Figs.~\ref{fig:onsite_dos}(c) and \ref{fig:onsite_dos}(d)]. The DOS therefore provides a direct spectroscopic signature of the robustness of the mixed-parity superconducting state.

	The observed selection of the NN $f$-wave channel can be understood from the
	$D_{3h}$ point-group classification of the triangular lattice. The NN
	$f$-wave basis function $f_{x(x^2-3y^2)}$ transforms as the fully symmetric
	$A_1'$ representation and is invariant under all point-group operations,
	making it compatible with the onsite-$s$-wave parent state. In contrast,
	the NNN $f^\prime$-wave basis function $f^\prime_{y(3x^2-y^2)}$ belongs to
	the $A_2'$ representation and is not fully symmetric, while the NN and NNN
	chiral $p$-wave channels belong to the two-dimensional $E'$ representation.
	This symmetry compatibility explains why the onsite-$s$ parent state
	selectively develops the NN $f$-wave component under an in-plane magnetic
	field.
	
	\subsection{Nearest-neighbor spin-singlet pairing}
	
	 Attractive interactions between nearest-neighbor sites in monolayer NbSe$_2$ naturally stabilize unconventional even-parity pairing states. Among the symmetry-allowed nearest-neighbor spin-singlet channels, the extended-$s$ and chiral-$d$ ($d_{x^2-y^2}\pm id_{xy}$) states are the leading candidates~\cite{Lu2018}. Since the field-induced equal-spin triplet correlations depend on the symmetry of the parent singlet condensate, we first determine the superconducting ground state by comparing the condensation energies of the competing pairing channels. We then investigate the mixed-parity superconducting state induced by the combined effects of Ising spin-orbit coupling and an in-plane Zeeman field.
	
	\subsubsection{\textbf{Thermodynamic stability and phase diagram of NN singlet pairing}}
	
	To determine the superconducting ground state, we compare the condensation energies of the self-consistent nearest-neighbor spin-singlet pairing states in monolayer NbSe$_2$.
	The corresponding grand-potential difference is given in Appendix~\ref{app:condensation_energy}. Throughout this section, the singlet interaction strength is fixed at $V_s=22.8$~meV, yielding a self-consistent superconducting gap of approximately $0.8$~meV at a realistic chemical potential, $\mu=-31.4$~meV.
	
	Figure~\ref{fig:NN_pairing}(a) compares the condensation energies of the chiral-$d$ and extended-$s$ pairing states as functions of chemical potential. At $\mu=-31.4$~meV, the chiral-$d$ state possesses the lower condensation energy and therefore constitutes the superconducting ground state. As $\mu$ increases, the free-energy difference between the two pairing states gradually decreases, and the condensation energies become nearly degenerate at $\mu\approx43.3$~meV. Beyond this crossover, the extended-$s$ state becomes energetically favorable. The inset enlarges the vicinity of the transition between the two competing pairing symmetries.
	
	\begin{figure*}[t]
		\centering
		
		\begin{minipage}[t]{0.24\textwidth}
			\centering
			\includegraphics[width=\linewidth]{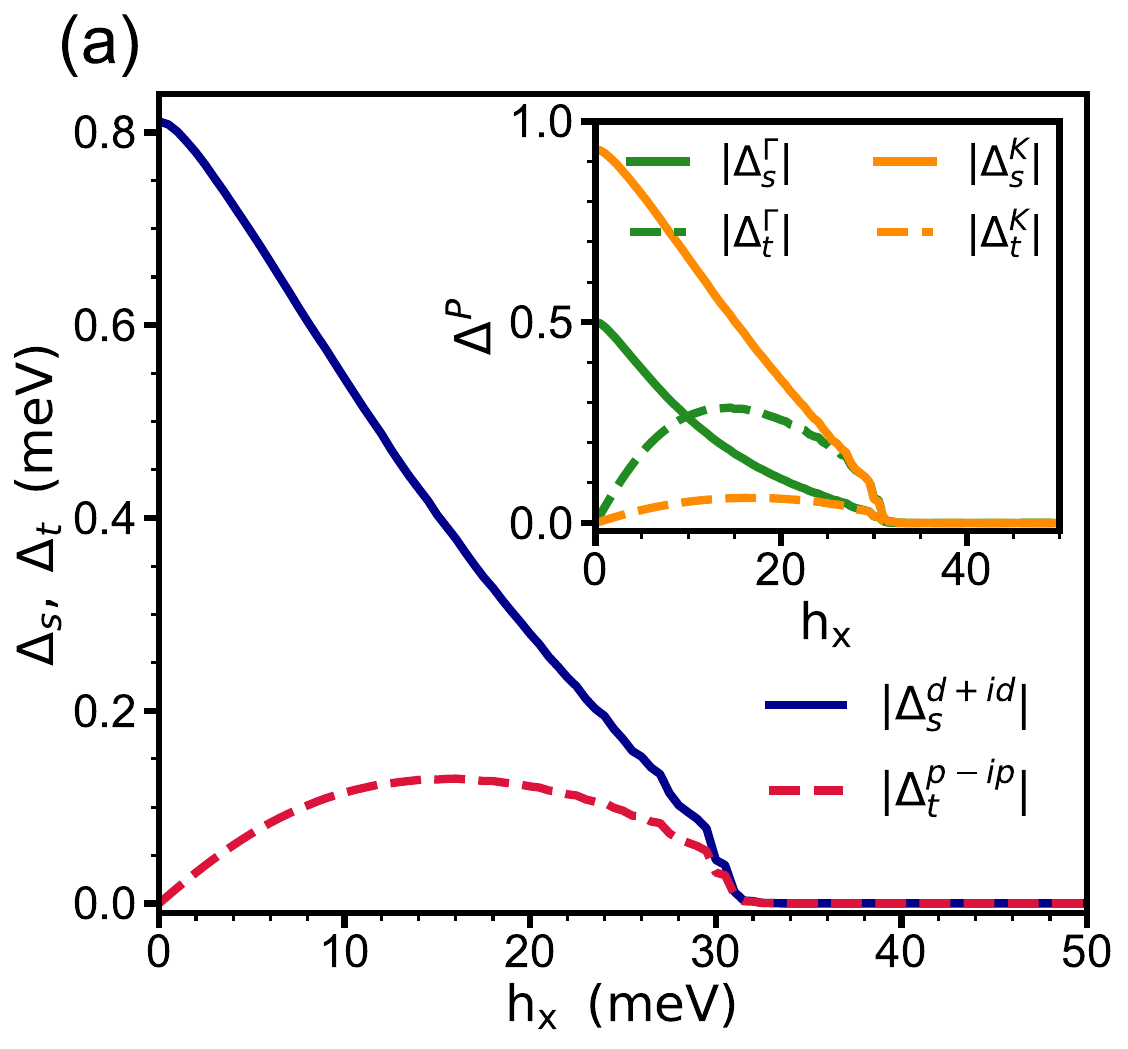}
		\end{minipage}
		\hfill
		\begin{minipage}[t]{0.24\textwidth}
			\centering
			\includegraphics[width=\linewidth]{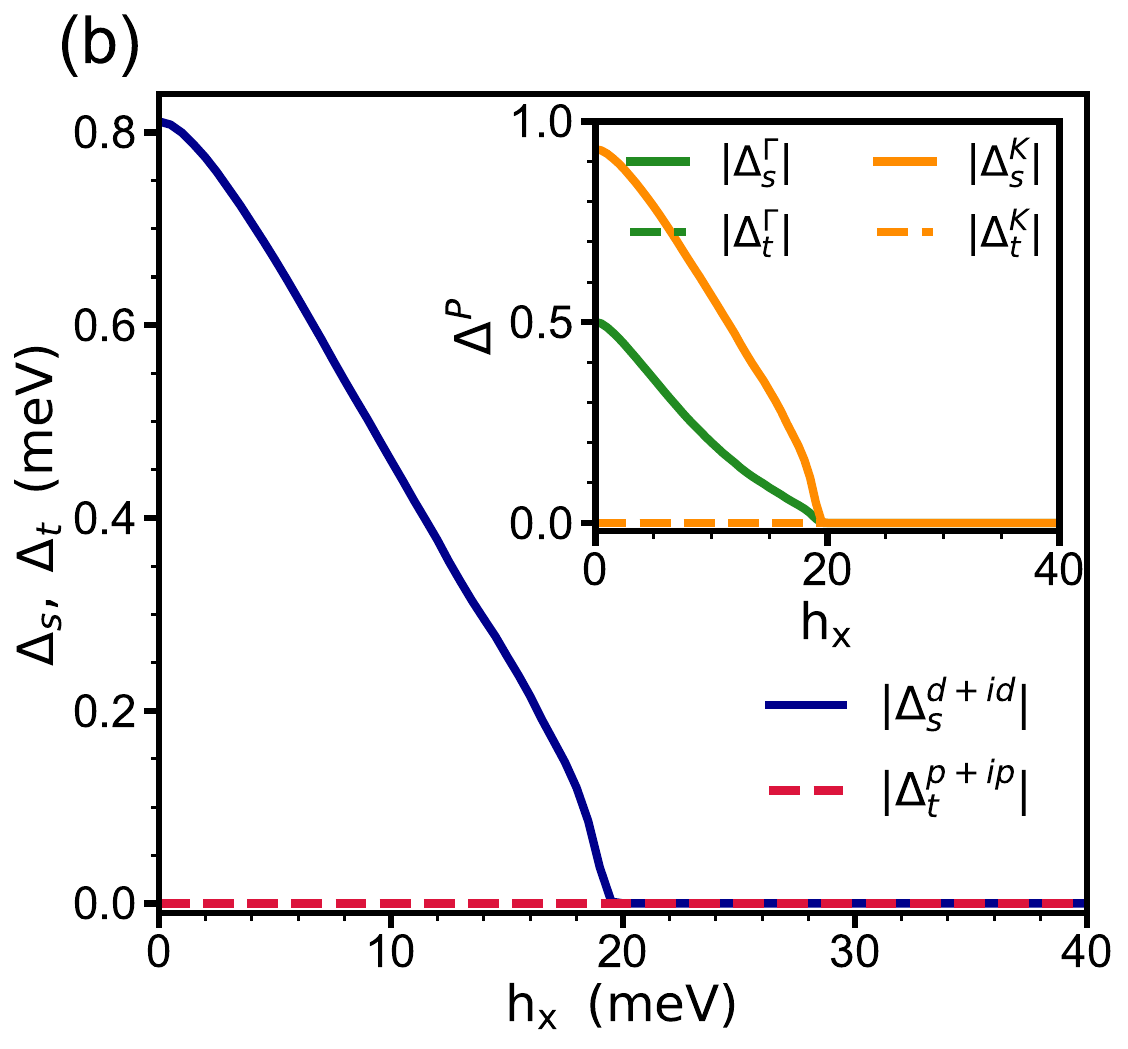}
		\end{minipage}
		\hfill
		\begin{minipage}[t]{0.24\textwidth}
			\centering
			\includegraphics[width=\linewidth]{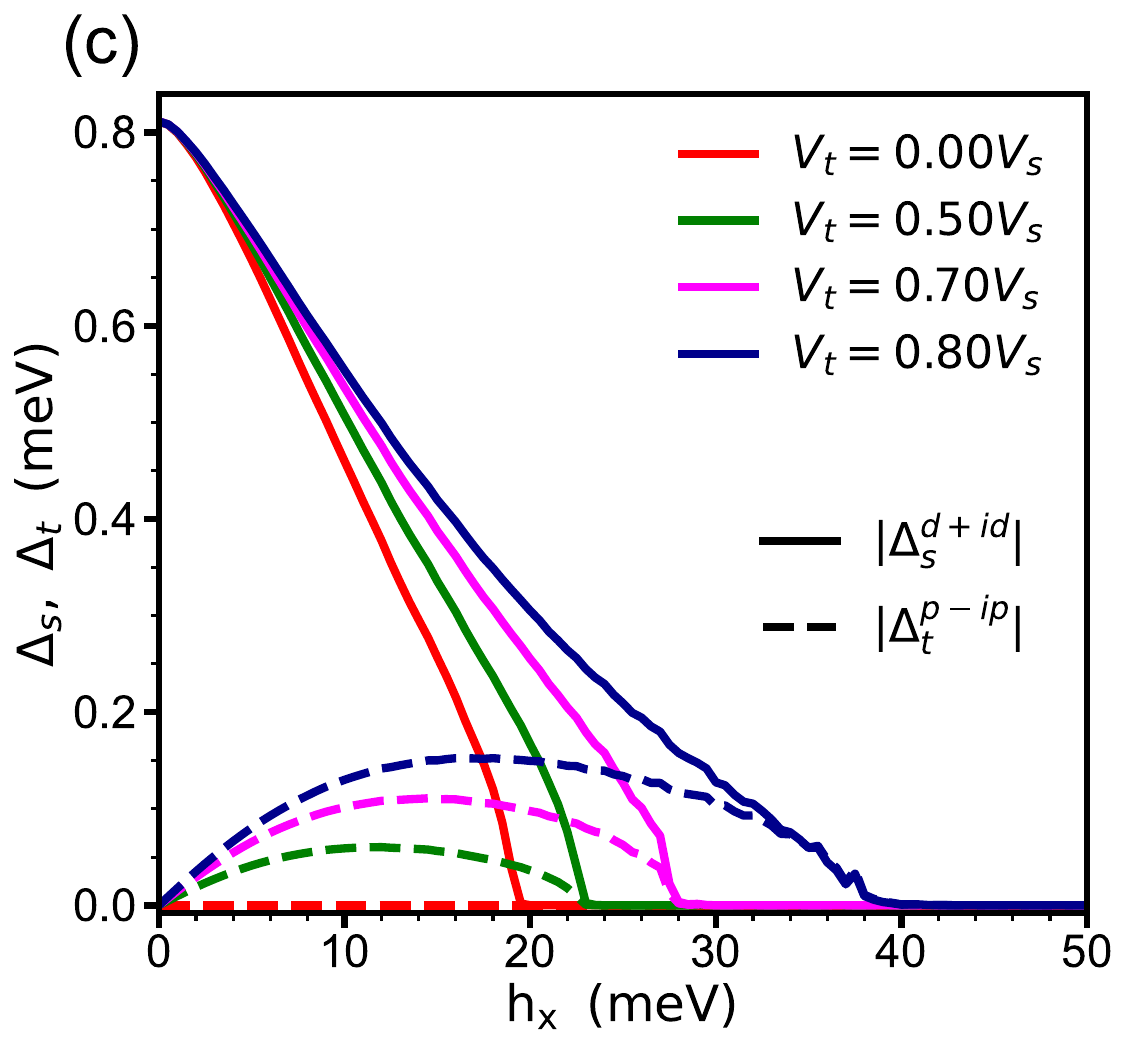}
		\end{minipage}
		\hfill
		\begin{minipage}[t]{0.24\textwidth}
			\centering
			\includegraphics[width=\linewidth]{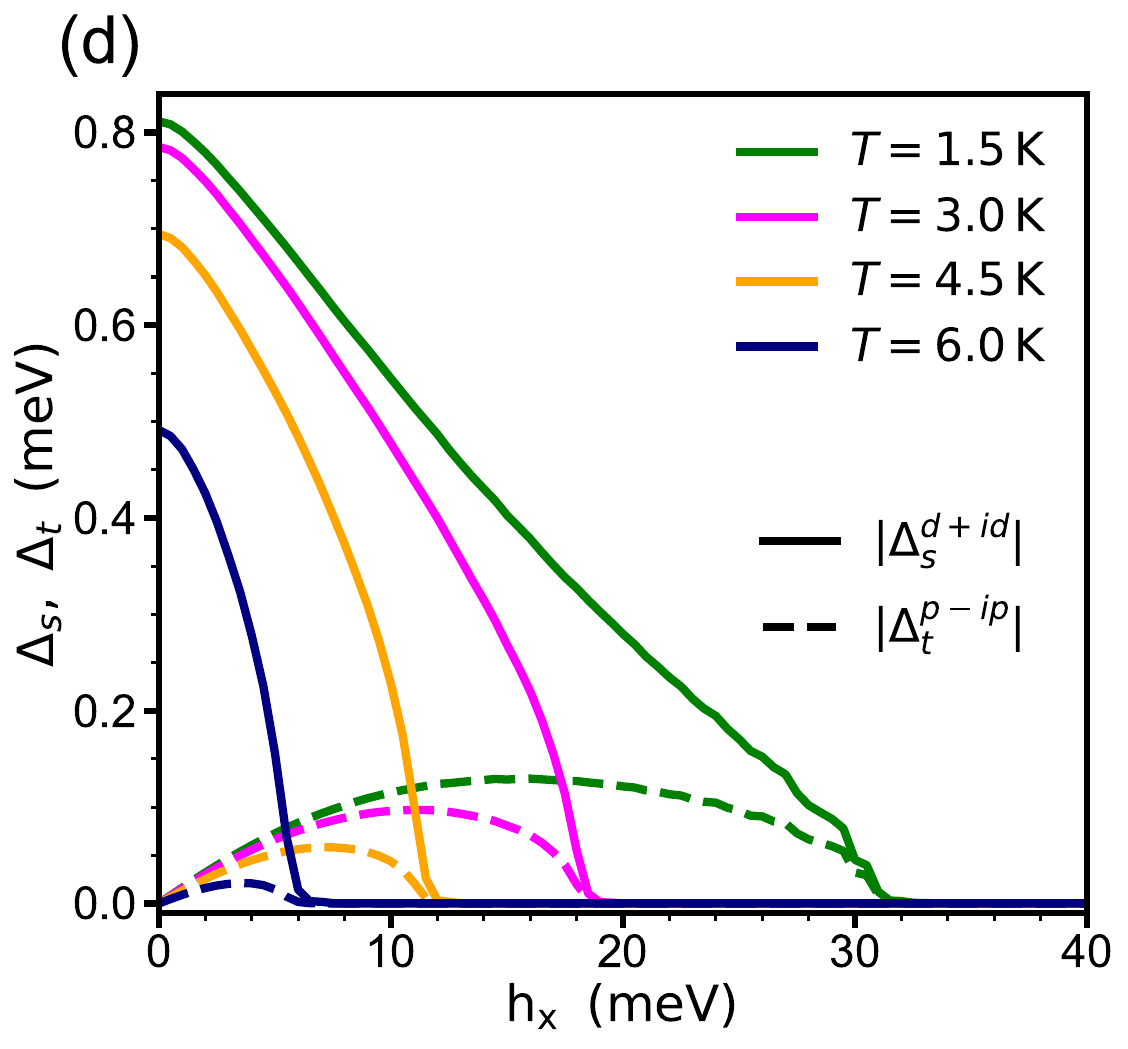}
		\end{minipage}
		
		\caption{
			Field-induced equal-spin triplet pairing for the parent chiral $d+id$ superconducting state.
			(a) Self-consistent singlet and induced chiral $p-ip$ triplet order parameters as functions of the in-plane magnetic field for $V_t=0.75V_s$. Insets show the corresponding $\Gamma$- and $K$-valley contributions.
			(b) Self-consistent solution for the $p+ip$ channel for $V_t=0.75V_s$, showing the absence of an induced triplet order parameter.
			(c) Evolution of the singlet and induced $p-ip$ order parameters for different triplet interaction strengths.
			(d) Magnetic-field dependence of the singlet $d+id$ and induced $p-ip$ order parameters at different temperatures for $V_t=0.75V_s$.
		}
		\label{fig:NN_triplet_order}
	\end{figure*}
	
	Using the pairing state with the lowest condensation energy at each chemical potential, we construct the superconducting phase diagram in the $(\mu,h_x)$ plane, shown in Fig.~\ref{fig:NN_pairing}(b).
	Two stable superconducting phases and the normal state arr shown in the parameter space. Around the  chemical potential reported for monolayer NbSe$_2$ (vertical dashed line), the chiral-$d$ state remains stable over a broad range of in-plane magnetic fields, whereas the extended-$s$ state is favored only at larger positive chemical potentials. Moreover, the chiral-$d$ phase exhibits a significantly higher critical field as compared to the onsite $s$-wave state, demonstrating that the chiral pairing state is more robust against in-plane magnetic fields.

	The temperature dependence of the self-consistent chiral-$d$ order parameter is shown in Fig.~\ref{fig:NN_pairing}(c), together with its momentum-resolved contributions from the regions near $\Gamma$, $K$, and $K'$ points. The superconducting gap decreases continuously with temperature and vanishes at $T_c\simeq7.4$~K, consistent with a second-order transition. Throughout the superconducting phase, 
		 the spectroscopic contribution to the order paremeter is dominant by the contributions surrounded by the valleys $K$ and $K'$. Although the electronic-state density within the Debye window is larger around the $\Gamma$ pocket, as established for the onsite-$s$ state, the NN chiral-$d$ form factor has substantially greater weight near the $K/K'$ valleys, resulting in a dominant pairing contribution from these regions. Thus, the NN chiral-$d$ state is predominantly associated with the $K/K'$ valleys.

	
	\subsubsection{\textbf{Field-induced equal-spin chiral-p triplet pairing}}
	
	Having established the chiral-$d$ state as the superconducting ground state, we next investigate the equal-spin triplet correlations induced by the combined effects of Ising spin-orbit coupling and an in-plane Zeeman field. To determine the symmetry of the field-induced triplet state associated with the parent $d+id$ superconducting phase, we solve the  singlet and triplet gap equations self-consistently for all symmetry-allowed odd-parity equal-spin pairing channels.
	
	Figures~\ref{fig:NN_triplet_order}(a) and (b) compare the self-consistent solutions for the two chiral triplet channels at $V_t=0.75V_s$. For a parent $d+id$ state, only the oppositely chiral $p-ip$ channel develops a finite equal-spin order parameter, whereas the the chiral-p order parameter  with same chirality remains zero. Self-consistent calculations for all symmetry-allowed odd-parity equal-spin pairing channels show that, besides the dominant NN chiral $p-ip$ state, only the NNN chiral $p-ip$ channel acquires a finite order parameter, whereas the remaining channels, including NN and NNN $f$-wave and NNN chiral $p+ip$ pairings, remain absent. The corresponding calculations are presented in
	Appendix~\ref{app:chiral_d_triplet_channels}. These results demonstrate that the symmetry of the induced triplet component is dictated by the symmetry of the parent $d+id$ condensate, selecting the oppositely chiral $p-ip$ state while excluding the state with same chirality~\cite{Akbar2024}. The inset of Fig.~\ref{fig:NN_triplet_order}(a) reveals a pronounced change in the momentum-space weight upon field-induced singlet--triplet conversion: the parent singlet is primarily concentrated in the $K/K'$ valleys, whereas the induced triplet contribution is predominantly concentrated around the $\Gamma$ pocket. This contrasting momentum dependence arises from the different form factors of the $d+id$ and $p-ip$ pairing channels, which favor different regions of the Brillouin zone.
	
	The symmetry selection is determined by the $D_{3h}$ point-group symmetry of monolayer NbSe$_2$. The NN and NNN $f$-wave basis functions belong to the $A_1'$ and $A_2'$ representations, respectively, and therefore cannot couple to the $d+id$ state, which belongs to $E'$. In contrast, the chiral $p$-wave states also belong to $E'$ and can therefore be induced by the parent $d+id$ state \cite{Hsu2017}. The chirality is determined by the threefold rotation symmetry $C_3$. Under $C_3$, $d+id$ acquires a phase $e^{i4\pi/3}$, while $p-ip$ acquires $e^{-i2\pi/3}=e^{i4\pi/3}$. Thus, $d+id$ and $p-ip$ transform in the same way under $C_3$, allowing them to couple. In contrast, $p+ip$ acquires a phase $e^{i2\pi/3}$ and cannot couple to $d+id$. This explains why the induced triplet component is $p-ip$, while the $f$-wave and $p+ip$ channels remain zero.
	
	\begin{figure}[b]
		\centering
		
		\begin{minipage}[t]{0.24\linewidth}
			\centering
			\includegraphics[height=3.4cm]{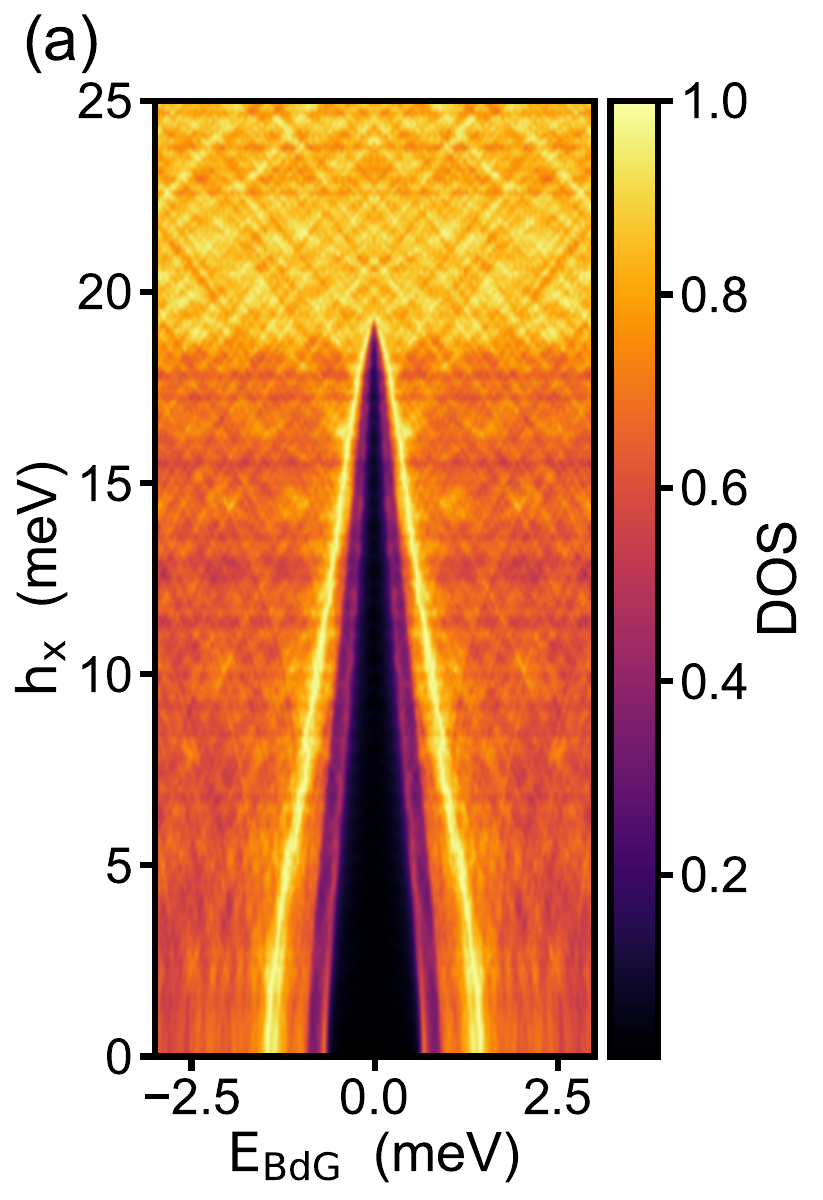}
		\end{minipage}
		\hfill
		\begin{minipage}[t]{0.24\linewidth}
			\centering
			\includegraphics[height=3.4cm]{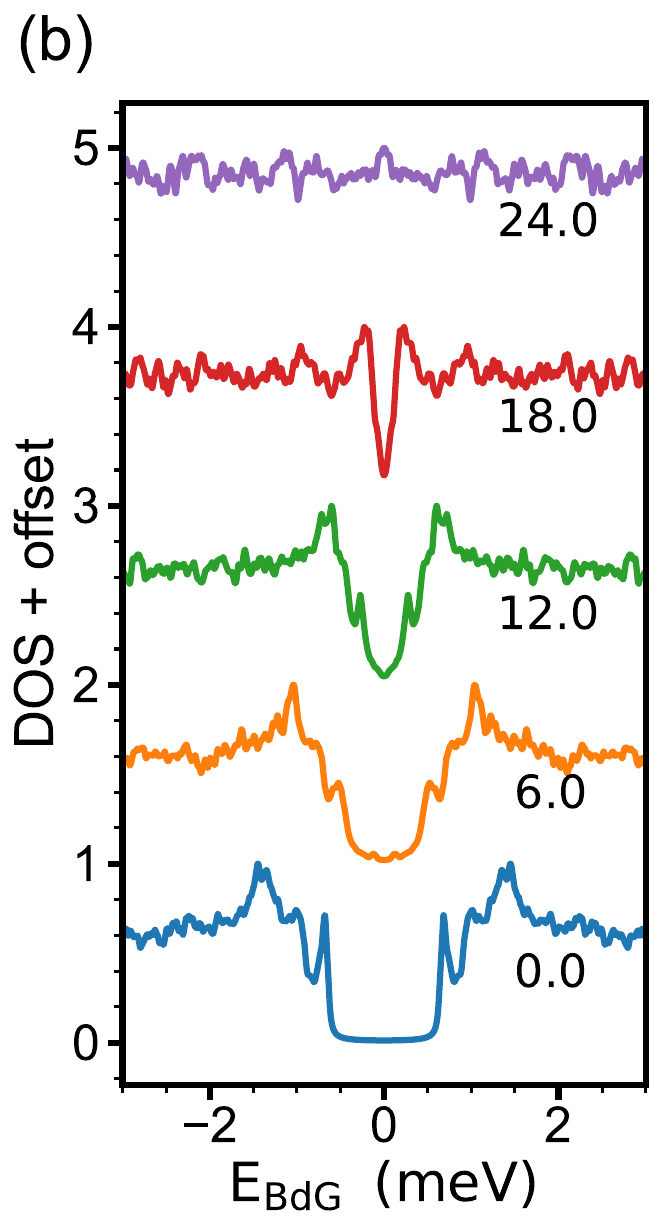}
		\end{minipage}
		\hfill
		\begin{minipage}[t]{0.24\linewidth}
			\centering
			\includegraphics[height=3.4cm]{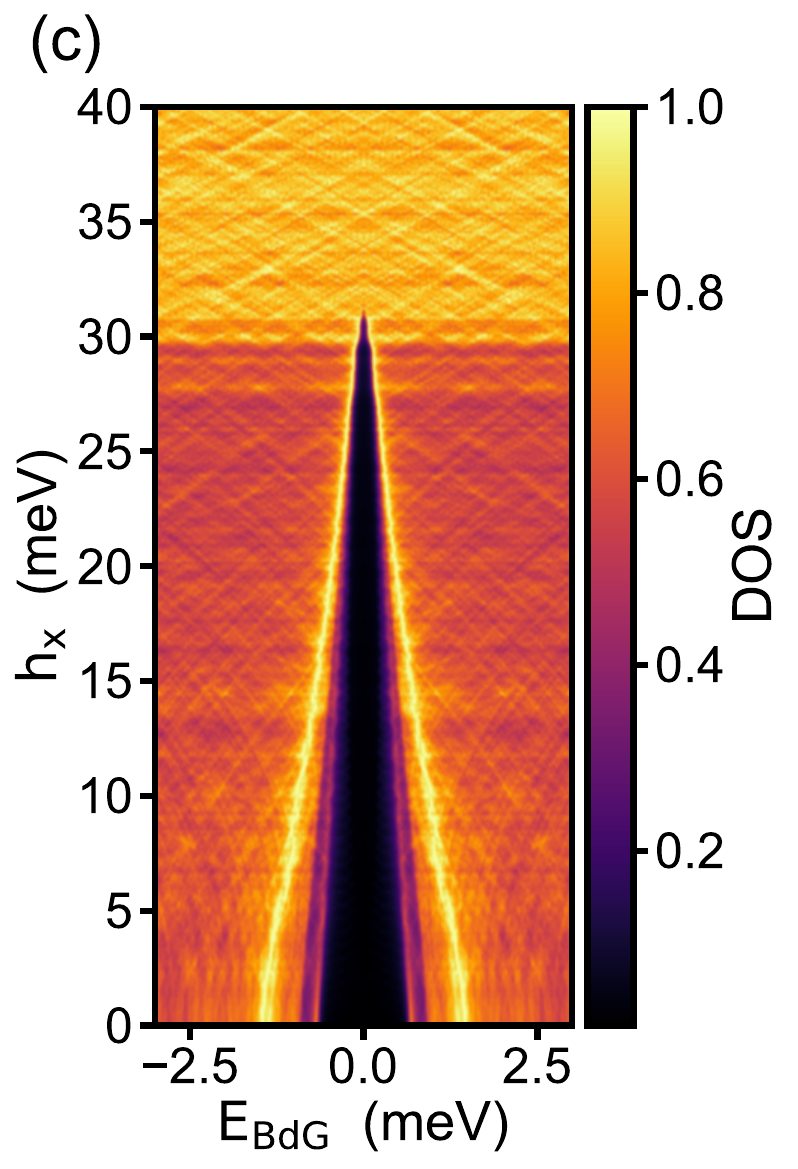}
		\end{minipage}
		\hfill
		\begin{minipage}[t]{0.24\linewidth}
			\centering
			\includegraphics[height=3.4cm]{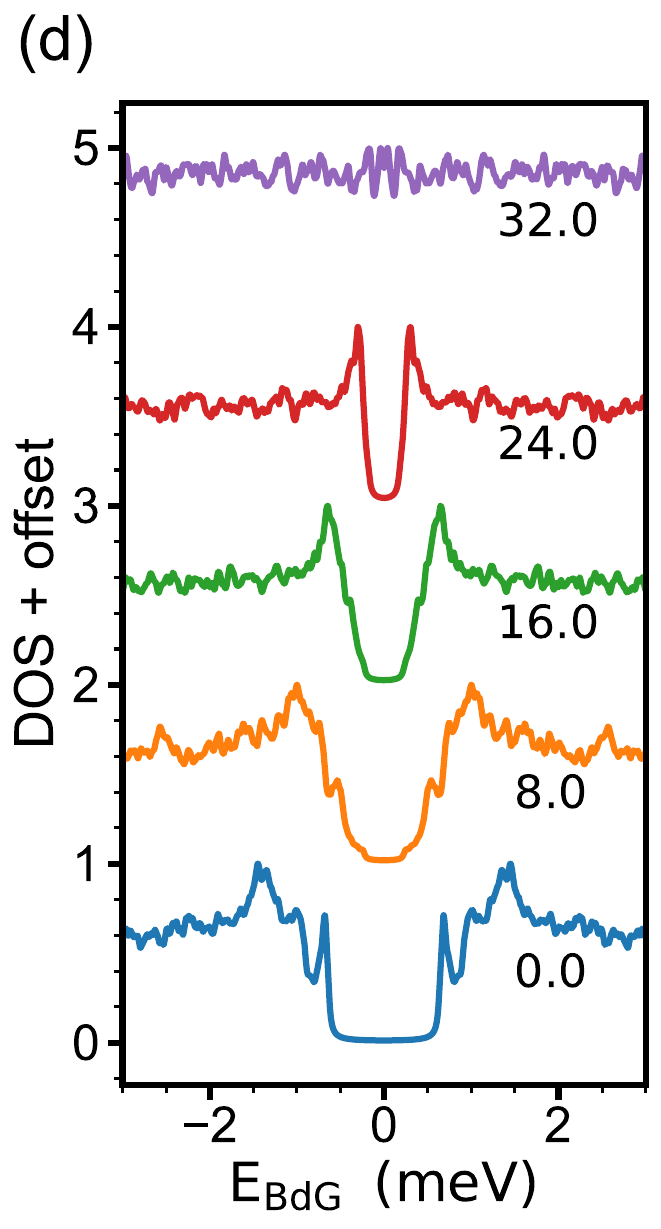}
		\end{minipage}
		
		\caption{
			Quasiparticle DOS of the chiral $d+id$ superconducting state in the presence of an in-plane Zeeman field.
			(a),(c) Field-resolved DOS heatmaps for the pure singlet state ($V_t=0$) and the mixed-parity state ($V_t=0.75V_s$), respectively.
			(b),(d) DOS spectra at selected values of $h_x$ (in meV) extracted from (a) and (c), respectively; successive curves are vertically offset for clarity.}
		\label{fig:NN_triplet_DOS}
	\end{figure}
	
	Having established the symmetry of the induced triplet state, we focus on the chiral $p-ip$ channel. Figure~\ref{fig:NN_triplet_order}(c) shows the evolution of the singlet and triplet order parameters for different intrinsic triplet interaction strengths. The induced chiral-$p$ component is absent at $V_t=0$, increases with the in-plane magnetic field through singlet-triplet conversion, reaches a maximum at intermediate fields, and vanishes together with the parent chiral $d+id$ order at the critical field. Increasing $V_t$ enhances the induced triplet order and shifts the critical field to higher values, demonstrating that intrinsic triplet interactions increase the robustness of the superconducting state.
	
	Figure~\ref{fig:NN_triplet_order}(d) shows the magnetic-field dependence of the singlet and induced triplet order parameters for different temperatures. As the temperature increases, both order parameters are progressively suppressed, leading to a monotonic reduction of the critical field. Nevertheless, the induced triplet component follows the evolution of the parent chiral $d+id$ order and vanishes simultaneously at the field-driven superconducting transition for each temperature, confirming that it remains induced by the parent singlet condensate.
	
	The quasiparticle density of states shown in Fig.~\ref{fig:NN_triplet_DOS}. Figures~\ref{fig:NN_triplet_DOS}(a) and (b) present the field evolution of the DOS for the parent chiral $d+id$ state ($V_t=0$). As the in-plane magnetic field increases, the superconducting gap gradually decreases and closes at the critical field, accompanied by the disappearance of the coherence peaks.
	
	The effect of the induced triplet pairing is shown in Figs.~\ref{fig:NN_triplet_DOS}(c) and (d) for $V_t=0.75V_s$. While the overall spectral evolution remains qualitatively similar, the superconducting gap persists to substantially higher magnetic fields, consistent with the enhanced critical field obtained from the self-consistent gap equations. The persistence of a finite excitation gap provides a spectroscopic signature of the enhanced robustness of the field-induced mixed-parity superconducting state.
	
	The emergence of a field-induced equal-spin chiral $p-ip$ state naturally raises the question of its robustness against additional spin-orbit interactions. We therefore investigate the influence of Rashba spin-orbit coupling on the superconducting phase diagram and the stability of the induced triplet state.
	
	\subsubsection{\textbf{Effect of Rashba spin-orbit coupling}}
	While superconductivity in pristine monolayer NbSe$_2$ is primarily governed by Ising
	spin-orbit coupling, Rashba spin-orbit coupling may arise in
	experimental devices due to substrate effects or electrostatic
	gating \cite{Shaffer2020,Cohen2024,Hanis2024}. We therefore include a
	nearest-neighbor Rashba term in the normal-state Hamiltonian
	$\eqref{eq:normal_ham}$,
	\begin{equation}
		H_R =
		i\lambda_R t_1
		\sum_{\langle ij\rangle_1}
		\sum_{\sigma\sigma'}
		c^\dagger_{i\sigma}
		\left[
		(\boldsymbol{\sigma}\times\mathbf{d}_{ij}^{\,1})
		\cdot\hat{\mathbf z}
		\right]_{\sigma\sigma'}
		c_{j\sigma'},
	\end{equation}
	where $\lambda_R$ is the dimensionless Rashba coupling,
	$\mathbf{d}_{ij}^{\,1}$ is the nearest-neighbor bond vector, and
	$\boldsymbol{\sigma}=(\sigma_x,\sigma_y,\sigma_z)$ denotes the Pauli
	spin matrices. Thus, the normal-state Hamiltonian becomes
	$H_0\rightarrow H_0+H_R$. In momentum space, this term takes the form
	\begin{equation}
		H_R(\mathbf{k})
		=
		\boldsymbol{\alpha}_R(\mathbf{k})
		\cdot\boldsymbol{\sigma}_{\parallel},
	\end{equation}
	where $\boldsymbol{\sigma}_{\parallel}=(\sigma_x,\sigma_y)$ and
	$\boldsymbol{\alpha}_R=(\alpha_{R,x},\alpha_{R,y})$ is the
	momentum-dependent Rashba field. For the triangular lattice,
	\begin{align}
		\alpha_{R,x} &=
		2\sqrt{3}\lambda_R t_1\sin\beta\cos\alpha,
		\nonumber\\
		\alpha_{R,y} &=
		-2\lambda_R t_1
		\left(\sin2\alpha+\sin\alpha\cos\beta\right),
	\end{align}
	with $\alpha=k_x/2$ and $\beta=\sqrt{3}k_y/2$.
	
	\begin{figure}[t]
		\centering
		\begin{minipage}[t]{0.49\linewidth}
			\centering
			\includegraphics[width=\linewidth]{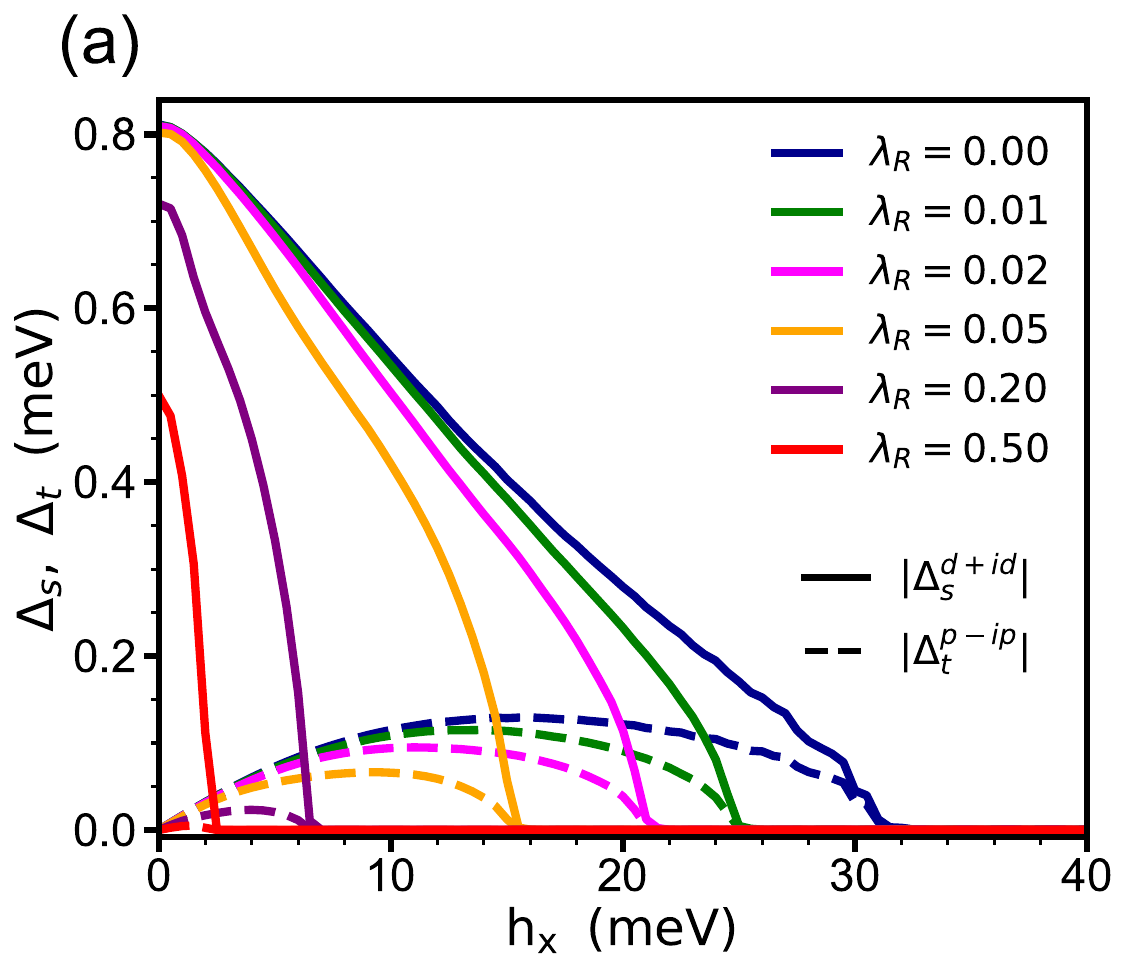}
		\end{minipage}
		\hfill
		\begin{minipage}[t]{0.49\linewidth}
			\centering
			\includegraphics[width=\linewidth]{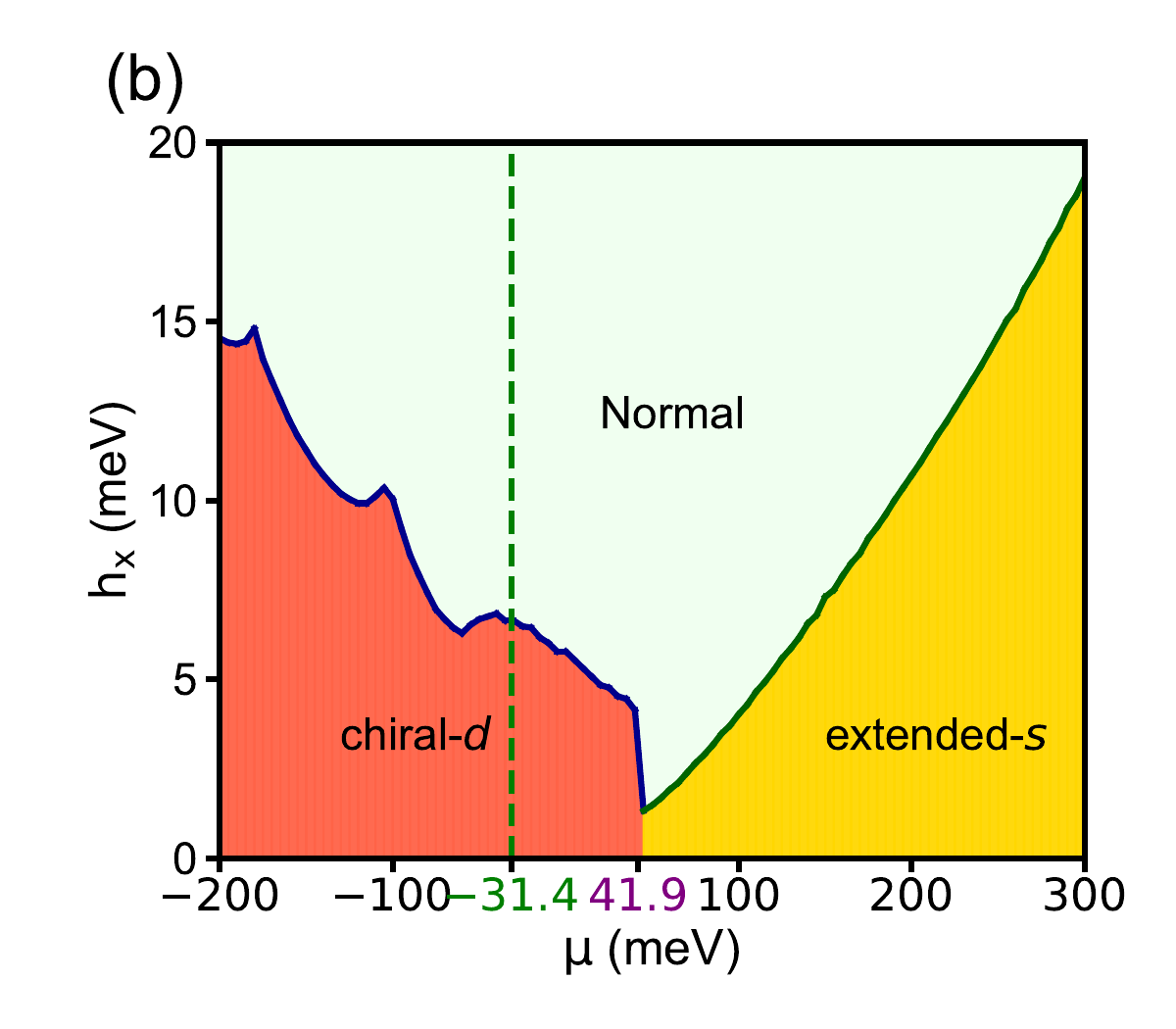}
		\end{minipage}
		\caption{
			(a) Magnetic-field dependence of the self-consistent chiral-$d$ singlet and induced chiral $p-ip$ triplet order parameters for different Rashba spin-orbit coupling strengths. Increasing $\lambda_R$ suppresses both order parameters and reduces the critical field.
			(b) Superconducting phase diagram in the $(h_x,\mu)$ plane for $\lambda_R=0.2$. The vertical dashed line denotes the physical chemical potential of monolayer NbSe$_2$.}
		\label{fig:rashba}
	\end{figure}
	
	The corresponding evolution of the singlet and induced triplet order parameters is shown in Fig.~\ref{fig:rashba}(a). Increasing the Rashba spin-orbit coupling suppresses both the parent chiral $d+id$ order and the induced chiral $p-ip$ triplet component, leading to a monotonic reduction of the critical field. Although the induced triplet pairing remains finite at moderate Rashba coupling, it is progressively suppressed with increasing $\lambda_R$.
	
	The resulting superconducting phase diagram for $\lambda_R=0.2$ is shown in Fig.~\ref{fig:rashba}(b). Compared with the pure Ising case [Fig.~\ref{fig:NN_pairing}(b)], the overall phase topology remains unchanged: the chiral-$d$ state remains the thermodynamically stable phase around the relevant  chemical potential of monolayer NbSe$_2$, whereas the extended-$s$ state is stabilized only at larger positive chemical potentials. However, the critical in-plane magnetic field is reduced throughout the phase diagram, with a more pronounced suppression of the chiral-$d$ phase, indicating that Rashba spin-orbit coupling weakens the Ising protection against Zeeman pair breaking.

	\subsection{Next-nearest-neighbor spin-singlet pairing}
	
\begin{figure}[t]
		\centering
		\begin{minipage}[t]{0.49\linewidth}
			\centering
			\includegraphics[width=\linewidth]{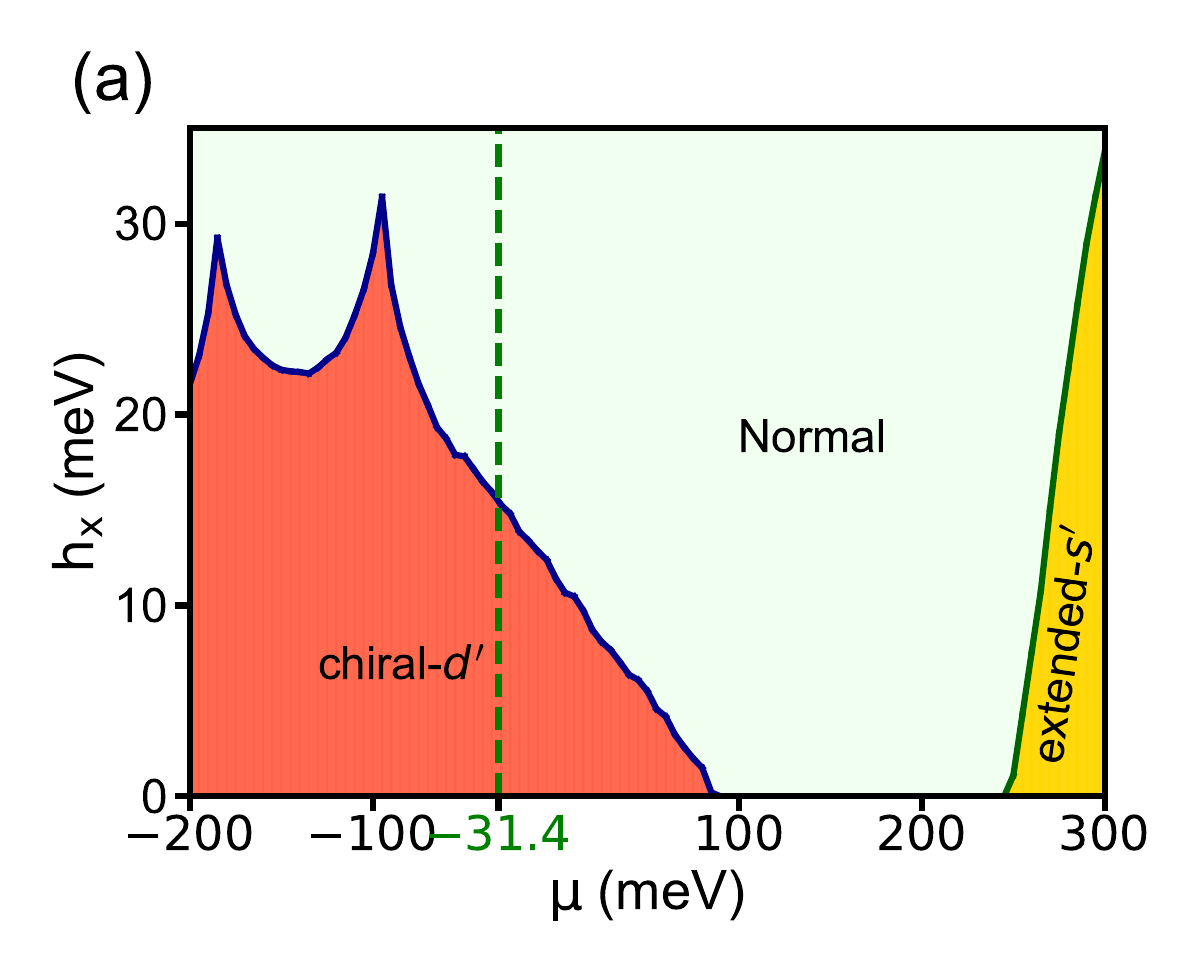}
		\end{minipage}
		\hfill
		\begin{minipage}[t]{0.49\linewidth}
			\centering
			\includegraphics[width=\linewidth]{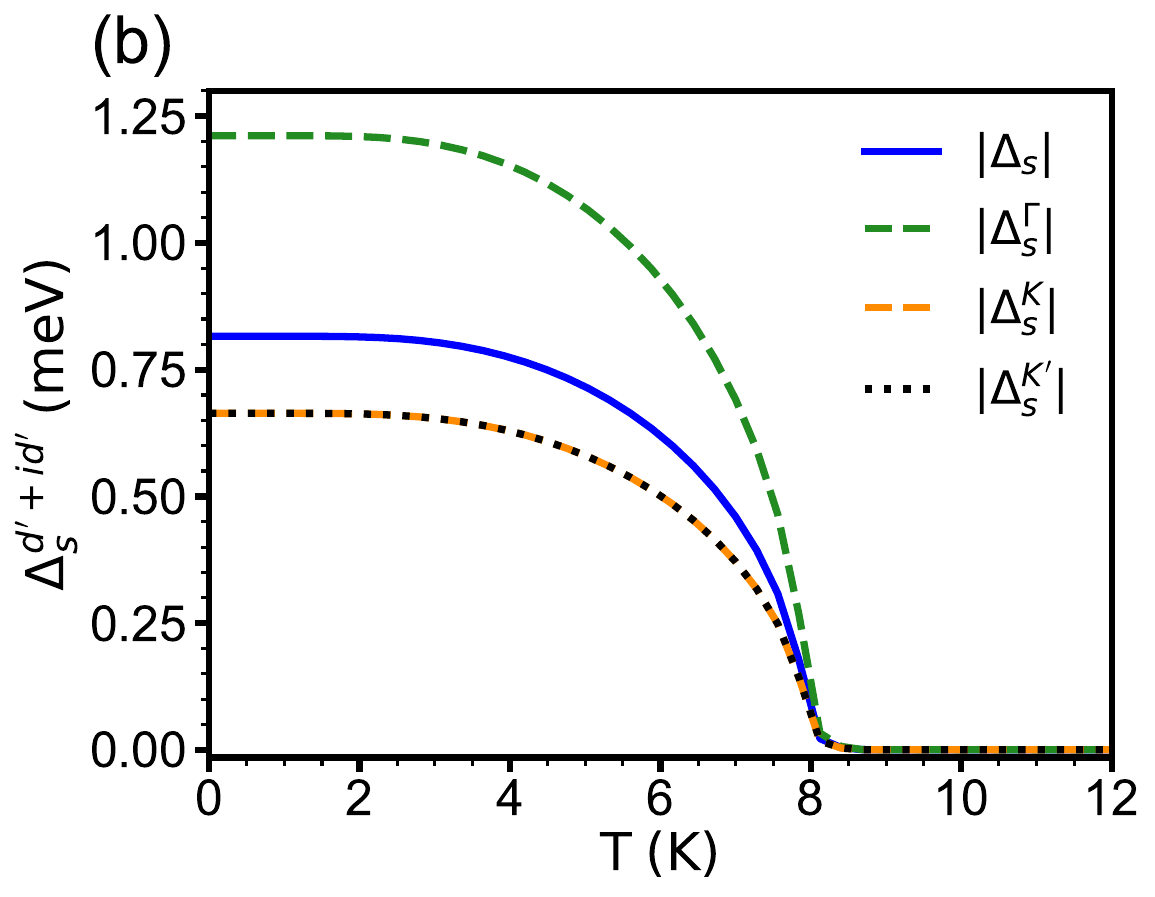}
		\end{minipage}
		
		\caption{
			NNN spin-singlet pairing: (a) superconducting phase diagram in the $(h_x,\mu)$ plane, showing that the chiral $d'+id'$ state is favored around the reported chemical potential $\mu=-31.4$ meV. (b) Temperature evolution of the $d'+id'$ state, showing the dominant contribution from the region near $\Gamma$ compared with the region near $K/K'$.
		}
		\label{fig:NNN_pairing}
	\end{figure}

	We next consider spin-singlet pairing between next-nearest-neighbor sites in monolayer NbSe$_2$. The symmetry-allowed NNN channels considered here are the extended-$s'$ and chiral-$d'$ states. Following the NN analysis, we determine the stable superconducting state and examine its response to an in-plane magnetic field. The corresponding phase diagram and temperature dependence of the chiral-$d'$ state are shown in Fig.~\ref{fig:NNN_pairing}.
	
	The phase diagram in Fig.~\ref{fig:NNN_pairing}(a) shows several differences from the NN case [Fig.~\ref{fig:NN_pairing}(b)]. Around the relevant chemical potential $\mu=-31.4$~meV, the chiral-$d'$ state remains stable over a broad field range, comparable to the NN chiral-$d$ state. A normal-state region appears at intermediate chemical potentials, where the fixed pairing strength $V_s=17.1$ meV is insufficient to sustain superconductivity. On the positive-$\mu$ side, the chiral-$d'$ state persists up to $\mu\simeq90$ meV, beyond which superconductivity is suppressed. The extended-$s'$ state emerges only at much larger positive chemical potentials, around $\mu\simeq250$ meV, and its critical field increases sharply with $\mu$. Thus, while the NNN chiral-$d'$ state shows similar field robustness near $\mu=-31.4$ meV, its phase diagram differs from the NN case on the positive-$\mu$ side.
	
	At $\mu=-31.4$ meV, the temperature dependence of the NNN chiral $d^\prime+id^\prime$ order parameter is shown in Fig.~\ref{fig:NNN_pairing}(b). The order parameter decreases continuously with temperature and vanishes at $T_c\simeq8.4$ K. This is slightly higher than the $T_c\simeq7.4$ K obtained for the NN $d+id$ state [Fig.~\ref{fig:NN_pairing}(c)]. 
	
	In contrast to the NN case, where the chiral-$d$ order is dominated by the
	$K/K'$ valleys, the NNN chiral $d^\prime+id^\prime$ state receives its
	largest contribution from the $\Gamma$ pocket. The $\Gamma$ contribution
	remains larger than that from the regions near $K/K'$ throughout the
	superconducting phase, with both contributions decreasing continuously
	toward $T_c$. This change in the relative high-symmetry contributions reflects the different momentum dependence of the NNN pairing form factor, which redistributes the superconducting weight between the $\Gamma$ and $K/K'$ Fermi-surface regions while preserving the same chiral-$d$ symmetry.

	\subsubsection*{\textbf{Field-induced equal-spin chiral-$p$ triplet pairing}}
	
	As in the NN case, the chiral-$d'$ states are favored at
	$\mu=-31.4$ meV for NNN pairing. We take the $d'+id'$ state as the
	parent state and examine the resulting field-induced equal-spin
	triplet correlations. Both the NN and NNN chiral $p-ip$ channels acquire finite triplet order parameters, with the NN component being dominant, whereas the remaining triplet
	channels remain zero. The NNN $p'-ip'$ component and the vanishing
	triplet channels are presented in Appendix~\ref{app:NNN_chiral_d'_triplet_channels}.
	
\begin{figure}[t]
		\centering
		\begin{minipage}[t]{0.49\columnwidth}
			\centering
			\includegraphics[width=\linewidth]{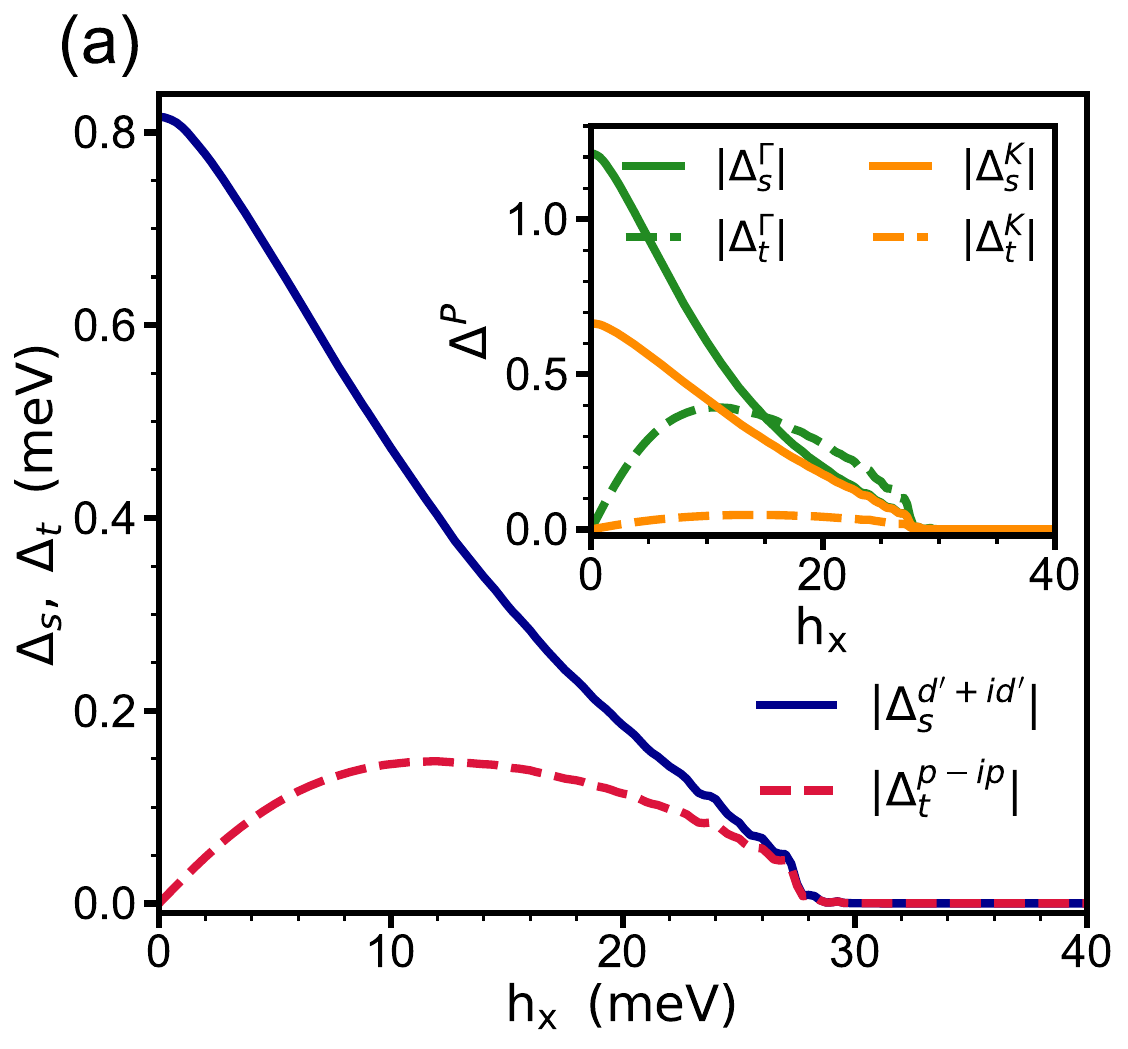}
		\end{minipage}
		\hfill
		\begin{minipage}[t]{0.49\columnwidth}
			\centering
			\includegraphics[width=\linewidth]{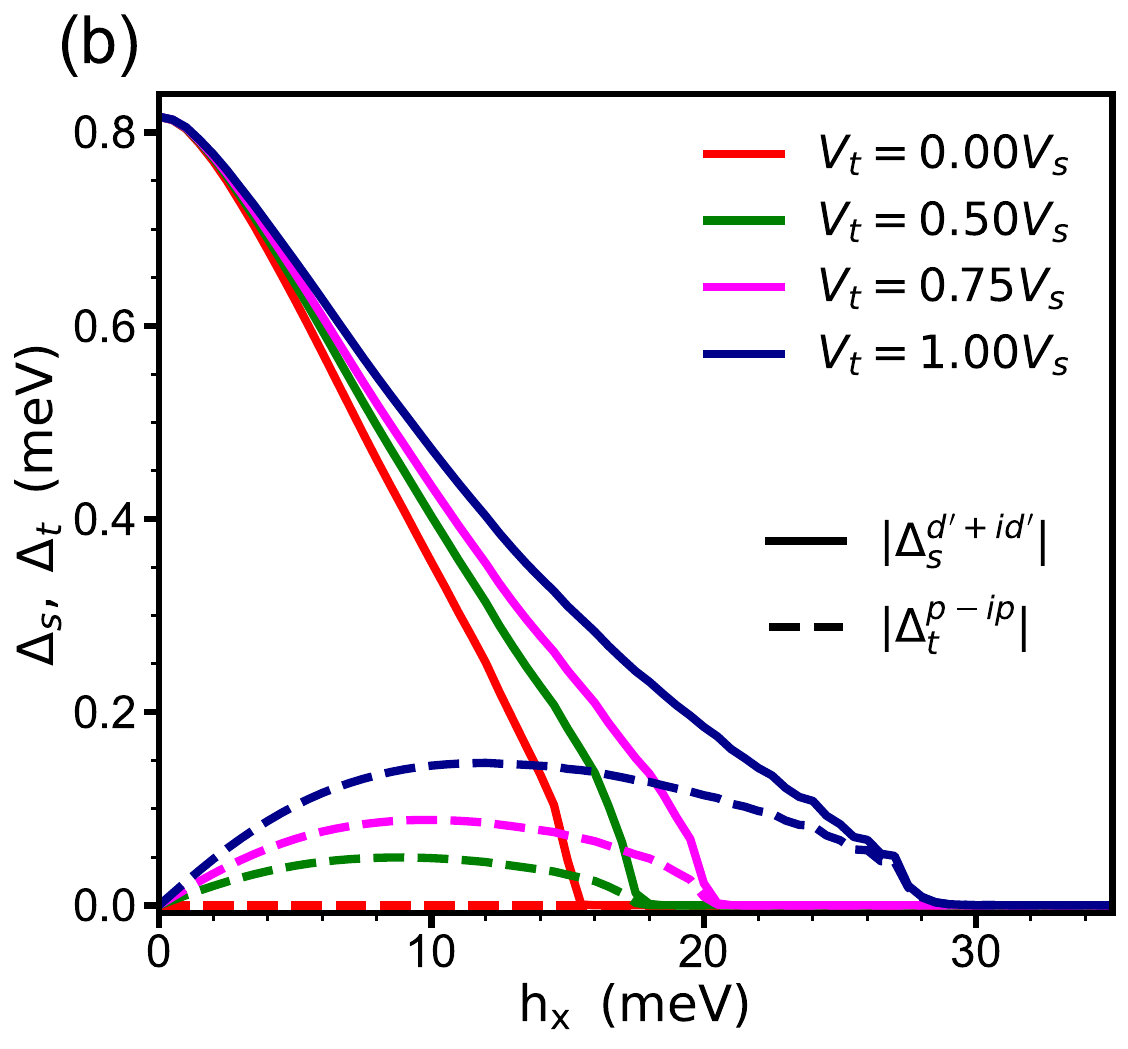}
		\end{minipage}
		\caption{(a) Self-consistent NNN chiral $d'+id'$ singlet and induced
			NN chiral $p-ip$ triplet order parameters as functions of the
			in-plane Zeeman field for $V_t=V_s$. The inset compares the
			contributions from the $\Gamma$ and $K/K'$ valleys. (b) Field
			dependence of the singlet and induced triplet order parameters
			for different triplet pairing strengths $V_t$.}
		\label{fig:NNN_triplet}
	\end{figure}

	Figure~\ref{fig:NNN_triplet}(a) shows the field dependence of the
	self-consistent NNN chiral $d'+id'$ singlet and induced NN chiral $p-ip$
	triplet order parameters for $V_t=V_s$. The singlet order parameter is
	distributed differently between the $\Gamma$ and $K/K'$ regions compared
	with the NN case shown in Fig.~\ref{fig:NN_triplet_order}(a). For NN
	pairing, the $K/K'$ contribution dominates, whereas for NNN pairing the
	$\Gamma$ contribution is larger. In contrast, the field-induced triplet
	component remains predominantly associated with the $\Gamma$ pocket, with
	only a comparatively small contribution from the $K/K'$ regions, similar
	to the NN case. Thus, changing the pairing range from NN to NNN primarily
	redistributes the singlet superconducting weight between the $\Gamma$ and
	$K/K'$ regions, while the induced triplet component retains its dominant
	$\Gamma$-pocket character. Both order parameters decrease with increasing
	field and vanish at the superconducting critical field.
	
	Figure~\ref{fig:NNN_triplet}(b) shows the evolution of the singlet and
	induced triplet order parameters for different values of $V_t$. As
	$V_t$ increases, the induced $p-ip$ component becomes progressively
	larger, while the superconducting state remains stable to higher
	in-plane fields. Compared with the NN case, the NNN state exhibits a
	slightly weaker field robustness despite having the same zero-field
	singlet gap. This difference reflects the distinct momentum structure
	of the NN and NNN pairing states and their corresponding response to
	the in-plane magnetic field.
	
	\section{\label{sec:V}Conclusions}
	
	We have studied the superconducting properties of
	monolayer NbSe$_2$, an intrinsic Ising superconductor, under an
	in-plane Zeeman field, with particular focus on the interplay between
	spin-singlet and field-induced equal-spin triplet pairing. We find that
	the thermodynamically favored singlet state depends on the pairing
	range, with distinct superconducting contributions from the regions near $\Gamma$, $K$, and $K'$ points in the Brillouin zone.
	 We further show that the field-induced triplet
	component is selectively determined by the symmetry of the parent
	singlet state and follows its evolution with magnetic field and
	temperature. The resulting mixed-parity states remain robust to
	in-plane fields, as reflected in their quasiparticle spectra.
	Additionally, we examine the effect of Rashba spin-orbit coupling
	associated with substrate coupling and electrostatic gating.

	\begin{table*}[hbt]
		\caption{Summary of the superconducting properties of the onsite, NN,
			and NNN spin-singlet parent states and their field-induced equal-spin
			triplet components, including the pairing strengths $V_s$ and $V_t$,
			critical temperature $T_c$, critical fields $h_x^c(V_t=0)$ and
			$h_x^c(V_t)$, and dominant high-symmetry-point (HSP) contributions.}
		\label{tab:pairing_summary}
		\begin{ruledtabular}
			\begin{tabular}{c c c c c c c c c c c c c c}
				Parent &
				Singlet &
				$V_s$ &
				$T_c$ &
				$h_x^c(V_t=0)$ &
				\multicolumn{1}{c}{Singlet HSP} &
				\multicolumn{4}{c}{NN triplet} &
				\multicolumn{4}{c}{NNN triplet} \\
				\noalign{\vskip 0.5mm}
				bond &
				pairing &
				(meV) &
				(K) &
				(meV) &
				&
				channel &
				HSP &
				$V_t/V_s$ &
				$h_x^c(V_t)$ &
				channel &
				HSP &
				$V_t/V_s$ &
				$h_x^c(V_t)$ \\
				
				\hline
				\noalign{\vskip 1.5mm}
				
				Onsite &
				$s$-wave &
				33.2 &
				5.8 &
				12.5 &
				$\Gamma > K$ &
				$f$-wave &
				$K \gtrsim \Gamma$ &
				0.75 &
				20.0 &
				-- &
				-- &
				-- &
				-- \\
				
				\noalign{\vskip 2mm}
				
				NN &
				$d+id$ &
				22.8 &
				7.4 &
				20.0 &
				$K \gg \Gamma$ &
				$p-ip$ &
				$\Gamma \gg K$ &
				0.75 &
				32.0 &
				$p'-ip'$ &
				$K > \Gamma$ &
				0.75 &
				28.0 \\
				
				\noalign{\vskip 2mm}
				
				NNN &
				$d'+id'$ &
				17.1 &
				8.4 &
				15.5 &
				$\Gamma \gg K$ &
				$p-ip$ &
				$\Gamma \gg K$ &
				1.00 &
				29.0 &
				$p'-ip'$ &
				$\Gamma$ ($K=0$) &
				1.00 &
				16.0 \\
				\noalign{\vskip 1mm}
				
			\end{tabular}
		\end{ruledtabular}
	\end{table*}
	
For onsite spin-singlet pairing, an in-plane Zeeman field induces an
	equal-spin NN $f$-wave triplet component through its interplay with the
	intrinsic Ising spin--orbit coupling. At low fields, the singlet order
	is dominated by the $\Gamma$ pocket, whereas the induced triplet
	component receives comparable contributions from the $\Gamma$ and
	$K/K'$ regions, with a slightly larger contribution from the latter.
	A finite triplet pairing interaction enhances the induced triplet
	component and increases the critical field.
	
For NN spin-singlet pairing, the chiral $d+id$ state is
	thermodynamically favored over the extended-$s$ state near
	$\mu=-31.4$~meV and is predominantly supported by the $K/K'$ valleys.
	The field-induced equal-spin triplet component has chiral $p-ip$
	symmetry, while the time-reversed $d-id$ state induces $p+ip$ pairing.
	Notably, the dominant NN chiral-$p$ component is concentrated around
	the $\Gamma$ pocket, in contrast to the $K/K'$-dominated singlet state,
	while the NNN chiral-$p$ component is subdominant. The induced triplet
	component enhances the field robustness of the mixed-parity state,
	whereas Rashba spin-orbit coupling suppresses both singlet and triplet
	components and reduces the critical field.
	
For NNN spin-singlet pairing, the chiral $d'+id'$ state is likewise
	thermodynamically favored near the physical chemical potential
	$\mu=-31.4$~meV, with the singlet contribution dominated by the
	$\Gamma$ pocket. The field-induced equal-spin triplet component retains
	the chirality correspondence, with $d'+id'$ ($d'-id'$) inducing
	$p-ip$ ($p+ip$) pairing. The dominant NN chiral-$p$ component remains
	$\Gamma$-centered, while the NNN component is subdominant. Thus, changing the pairing range shifts the dominant singlet contribution from $K/K'$ to $\Gamma$, while the induced triplet component remains predominantly $\Gamma$-centered and retains the chirality correspondence with the parent singlet state. A summary of the resulting properties is given in
	Table~\ref{tab:pairing_summary}.
	
	In summary, our results establish a direct connection between the
	momentum structure of the singlet parent state, the symmetry of the
	induced equal-spin triplet component, and the magnetic-field robustness
	of superconductivity in monolayer NbSe$_2$. In particular, the
	chirality locking between chiral-$d$ singlet and chiral-$p$ triplet
	pairing provides a symmetry-based mechanism for field-controlled
	mixed-parity superconductivity. These results highlight the possibility
	of probing field-induced equal-spin triplet correlations through
	momentum- and energy-resolved spectroscopic measurements and provide
	insight into the interplay between pairing symmetry and spin-orbit
	coupling in two-dimensional Ising superconductors.

	\section*{Acknowledgments}
	S.B. thanks Grigorii Bobkov for helpful initial comments regarding the modeling and numerical treatment of monolayer NbSe$_2$.


	\appendix
	
	
	\section{Triplet channels for onsite $s$-wave spin-singlet state}
	\label{app:onsite_s_triplet_channels}
	
\begin{figure}[hbt]
		\centering
		
		\begin{minipage}[t]{0.56\linewidth}
			\centering
			\includegraphics[width=\linewidth]{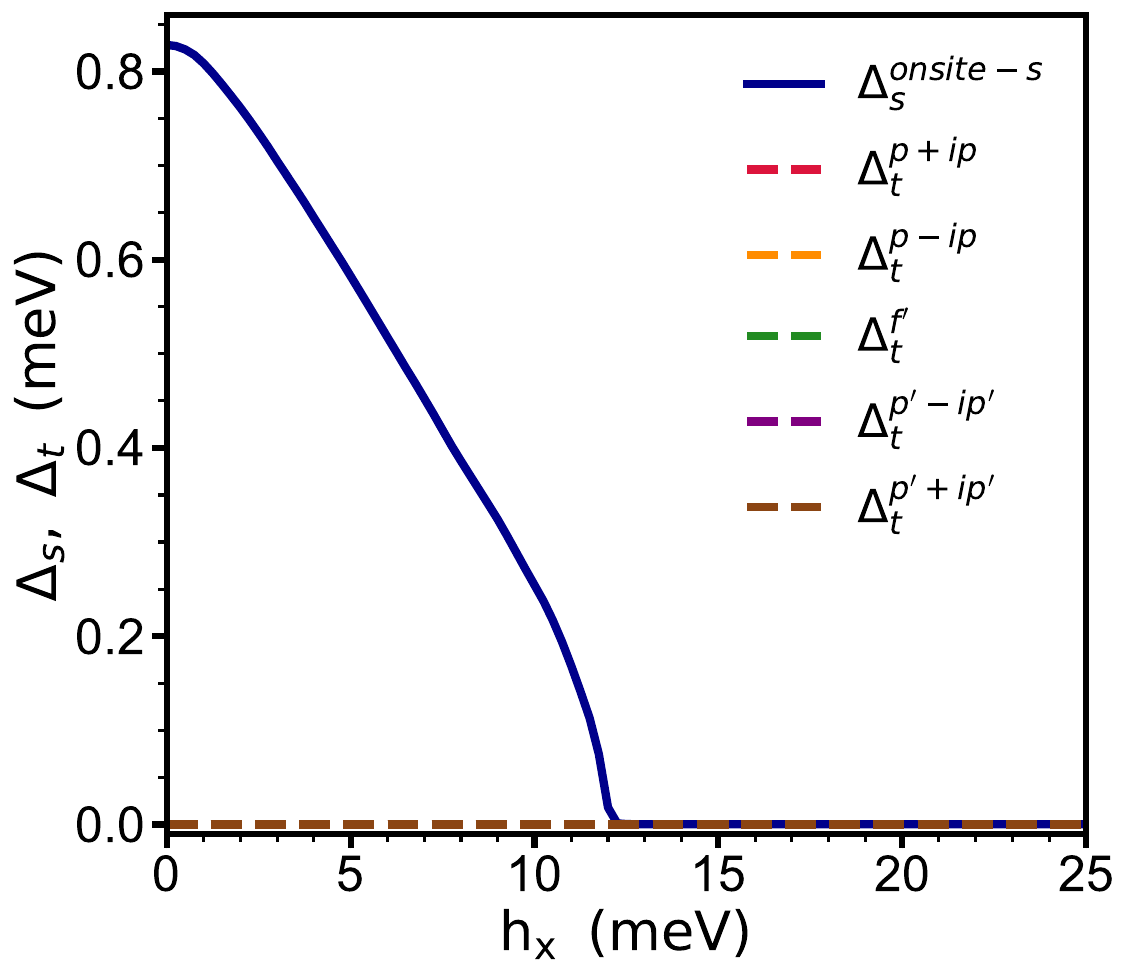}
		\end{minipage}

		\caption{ Field dependence of the self-consistent onsite-$s$-wave singlet and candidate equal-spin triplet order parameters, showing the absence of a finite triplet component in these channels.}
		\label{fig:other_triplet_channels}
	\end{figure}

	Starting from the onsite $s$-wave spin-singlet state, we solve the
	coupled singlet-triplet BdG equations independently for all
	symmetry-allowed NN and NNN triplet channels. As shown in
	Fig.~\ref{fig:other_triplet_channels}, the NN chiral
	$p\pm ip$, NNN $f'$-wave,  and chiral $p'\pm ip'$ channels remain zero within
	numerical accuracy throughout the superconducting phase. Thus, among the candidate channels, only the NN f-wave component is induced by the in-plane Zeeman field, as shown in Fig.~\ref{fig:onsite_fwave}(a).
	
	\section{Condensation energy of the pairing states}
	\label{app:condensation_energy}
	
	To determine the thermodynamically stable NN or NNN
	spin-singlet pairing state, we compare the grand potentials of the
	self-consistent chiral-$d$ and extended-$s$ solutions. Since this
	comparison is performed in the absence of an in-plane magnetic field
	and equal-spin triplet interaction, we set $h_x=0$ and $V_t=0$.
	Consequently, only the spin-singlet pairing component remains in the pairing matrix in Eq.~\eqref{eq:pairing_matrix}.
	
	\begin{figure*}[t]
		\centering
		
		\begin{minipage}[t]{0.28\linewidth}
			\centering
			\includegraphics[width=\linewidth]
			{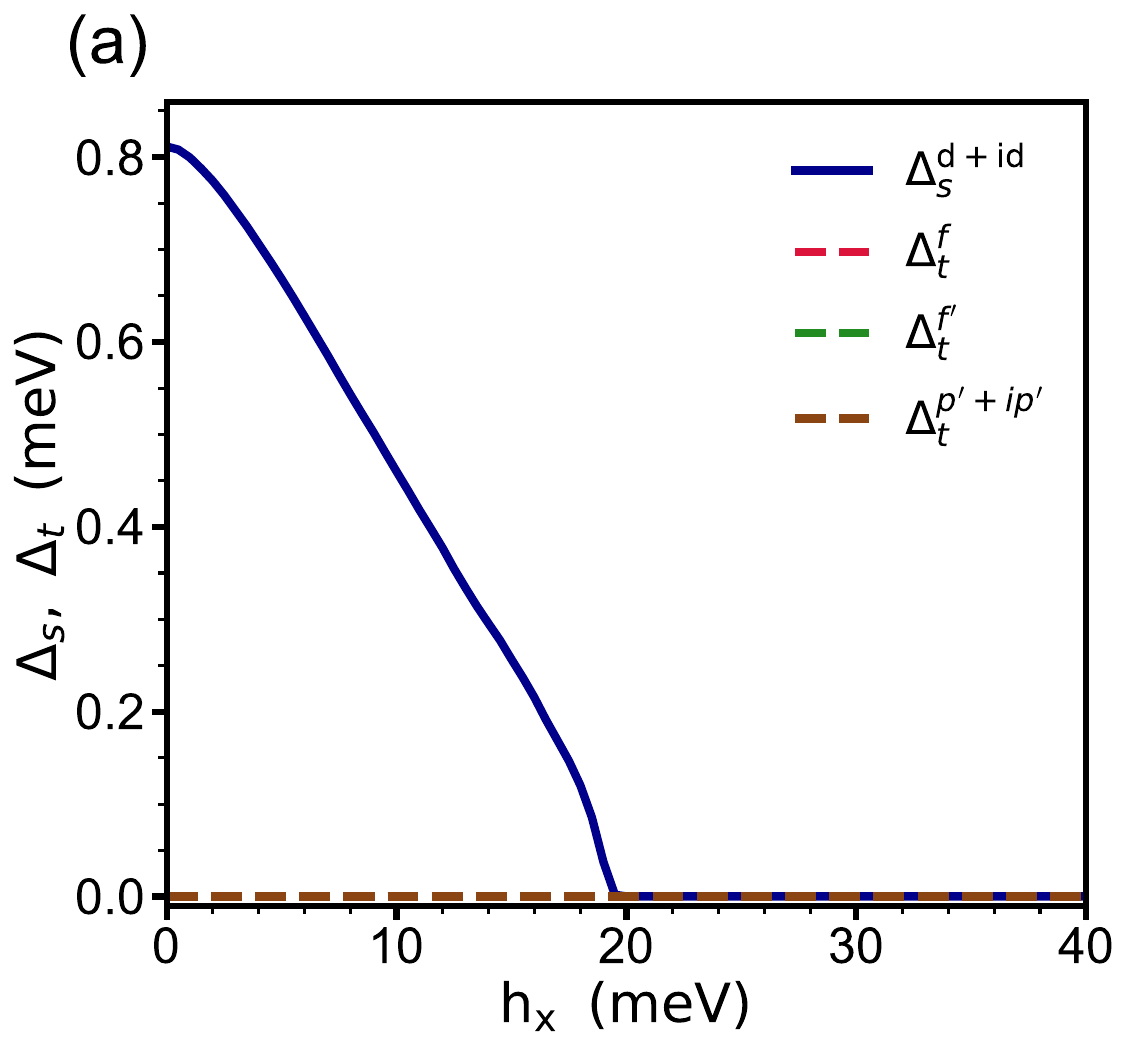}
		\end{minipage}
		\hfill
		\begin{minipage}[t]{0.28\linewidth}
			\centering
			\includegraphics[width=\linewidth]
			{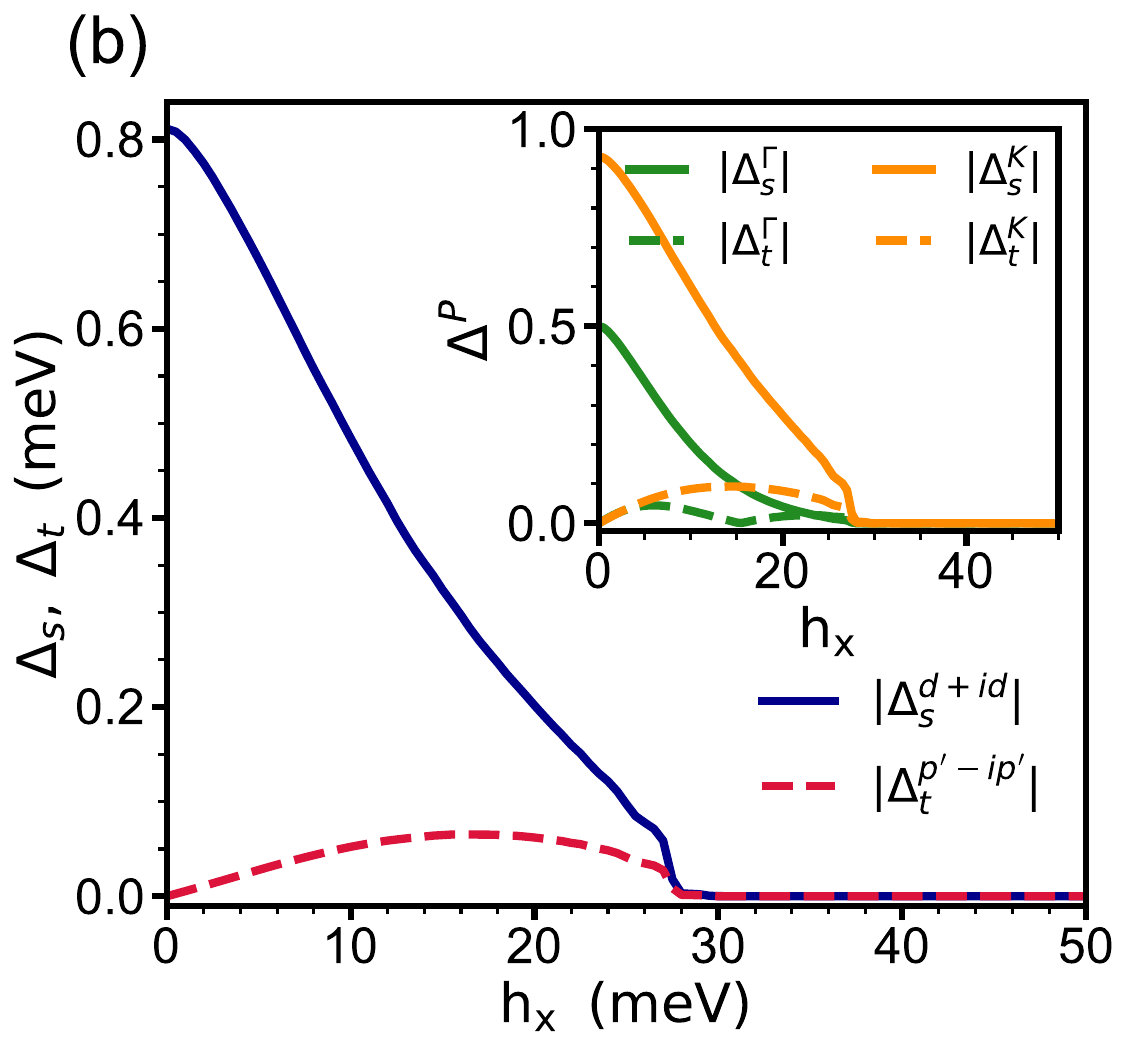}
		\end{minipage}
		\hfill
		\begin{minipage}[t]{0.28\linewidth}
			\centering
			\includegraphics[width=\linewidth]
			{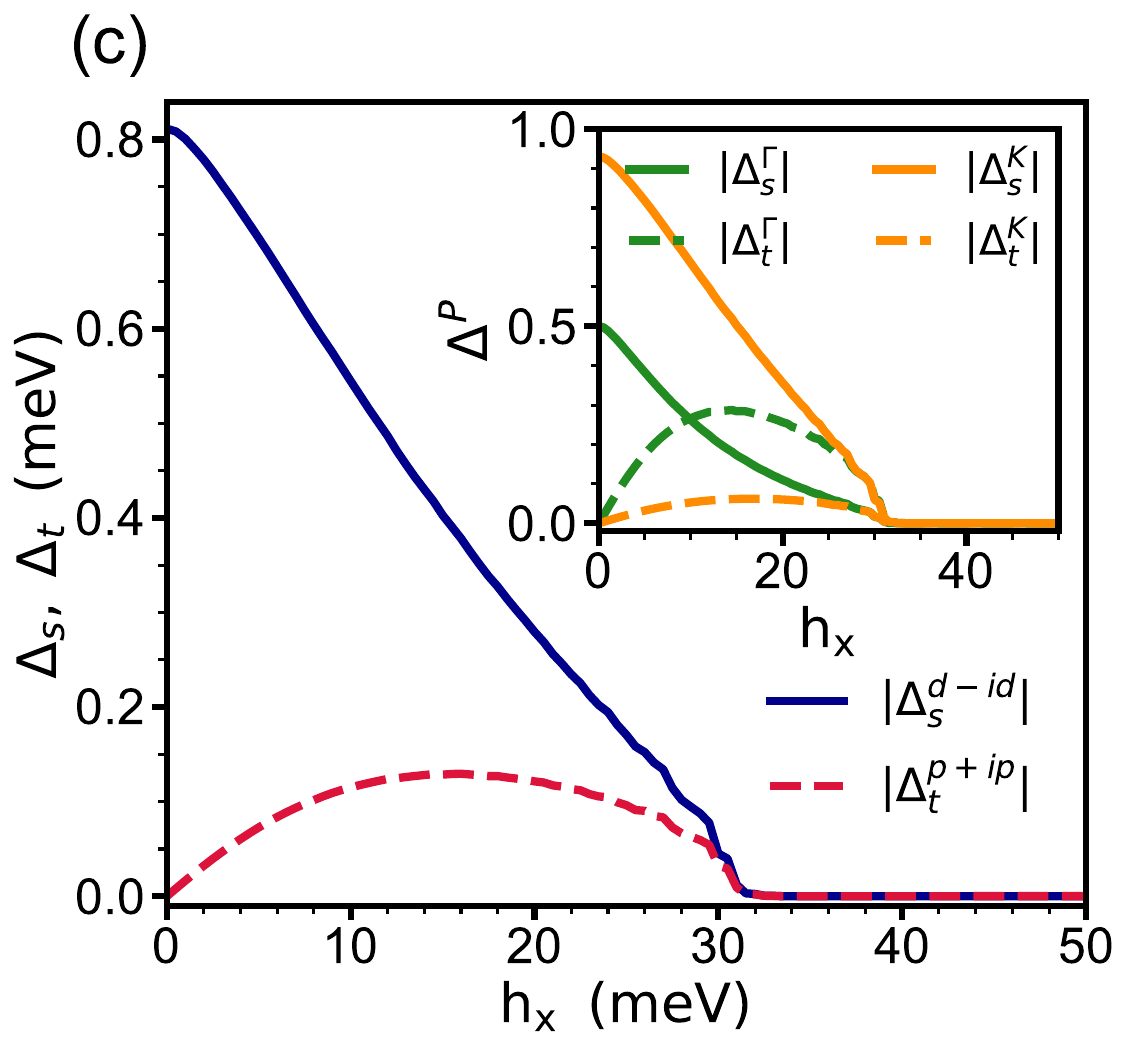}
		\end{minipage}
		
		\caption{
			Symmetry selection of the field-induced equal-spin triplet channels. For the $d+id$ parent state, the NN $f$-wave, NNN $f^\prime$-wave, and NNN chiral $p^\prime+ip^\prime$ components remain absent (a), whereas the NNN chiral $p^\prime-ip^\prime$ component is finite (b). Reversing the chirality of the parent state to $d-id$ correspondingly selects the NN chiral $p+ip$ channel (c). Insets show the $\Gamma$- and $K/K'$-resolved contributions.
		}
		\label{fig:chiral_d_all_triplet_symmetry}
	\end{figure*}
	
	For a self-consistent solution of $\Delta_s$, the corresponding
	mean-field grand potential is obtained by diagonalizing the BdG Hamiltonian in Eq.~\eqref{eq:BdG_Hamiltonian}. Denoting the positive BdG eigenvalues of the
	superconducting and normal-state Hamiltonians by
	$E_{n}^{\rm sc}(\mathbf{k})$ and $E_{n}^{\rm N}(\mathbf{k})$,
	respectively, the superconducting-normal-state grand-potential
	difference is
	\begin{equation}
		\begin{aligned}
			\Delta\Omega
			&\equiv \Omega_{\rm sc}-\Omega_{\rm N} \\
			&= \frac{|\Delta_s|^2}{V_s}
			-\frac{1}{\beta}
			\sum_{\mathbf{k}\in \mathbf{k}_D}
			\sum_{E_n>0}
			\Bigg[
			\ln\left(
			2\cosh\frac{\beta E_{n}^{\rm sc}(\mathbf{k})}{2}
			\right) \\
			&\hspace{3.2cm}
			-\ln\left(
			2\cosh\frac{\beta E_{n}^{\rm N}(\mathbf{k})}{2}
			\right)
			\Bigg].
		\end{aligned}
		\label{eq:deltaOmega_finiteT}
	\end{equation}
	where $\beta=(k_B T)^{-1}$ and $D$ denotes the Debye shell
	defined by $|E_n(\mathbf{k}_D)| \leq \omega_D$.The subtraction of the normal-state contribution removes the $\Delta_s$-independent part of the grand potential.
	
	For numerical calculations, we normalize the grand-potential
	difference by the number of momentum points in the Debye shell,
	$N_D$, and define the condensation energy density as
	\begin{equation}
		\delta\Omega
		=
		\frac{\Delta\Omega}{N_D} .
		\label{eq:condensation_energy}
	\end{equation}
	A negative value of $\delta\Omega$ indicates that the superconducting
	state is energetically favorable relative to the normal state.
	
	
	\section{Triplet channels for the NN chiral-$d$ spin-singlet state}
	\label{app:chiral_d_triplet_channels}

	For the $d+id$ parent state, the NN oppositely chiral $p-ip$ channel is the
	dominant finite triplet component, while the  NN channel with same chirality,
	$p+ip$, remains zero, as shown in Fig.~\ref{fig:NN_triplet_order}(a)
	and (b), respectively. To further examine the symmetry selection, we
	solve the self-consistent gap equations for the remaining odd-parity
	channels. As shown in Fig.~\ref{fig:chiral_d_all_triplet_symmetry}(a) and (b), the NN $f$-wave, NNN $f'$-wave and $p^\prime+ip^\prime$ channels do not develop a triplet component,
	whereas the NNN $p^\prime-ip^\prime$ channel acquires a finite component.
	Thus, for nonlocal triplet pairing, the $d+id$ singlet state selectively
	induces the $p-ip$ triplet component, with opposite chirality to the
	parent state.

%
%
%
	
	As an additional check, reversing the chirality of the parent
	superconducting state reverses the chirality of the induced triplet
	component. For a $d-id$ parent state, the $p+ip$ channel acquires a
	finite equal-spin triplet order parameter, as shown in
	Fig.~\ref{fig:chiral_d_all_triplet_symmetry}(c), while the state with same
	chirality is suppressed. Thus,
	\begin{equation}
		d+id \;\longrightarrow\; p-ip,
		\qquad
		d-id \;\longrightarrow\; p+ip.
	\end{equation}
	This correspondence demonstrates that the chirality of the induced
	triplet component is determined by that of the parent singlet state.

	
	\section{Triplet channels for the NNN chiral-$d'$ spin-singlet state}
	\label{app:NNN_chiral_d'_triplet_channels}
	
\begin{figure*}[hbt]
		\centering
		
		\begin{minipage}[t]{0.28\linewidth}
			\centering
			\includegraphics[width=\linewidth]
			{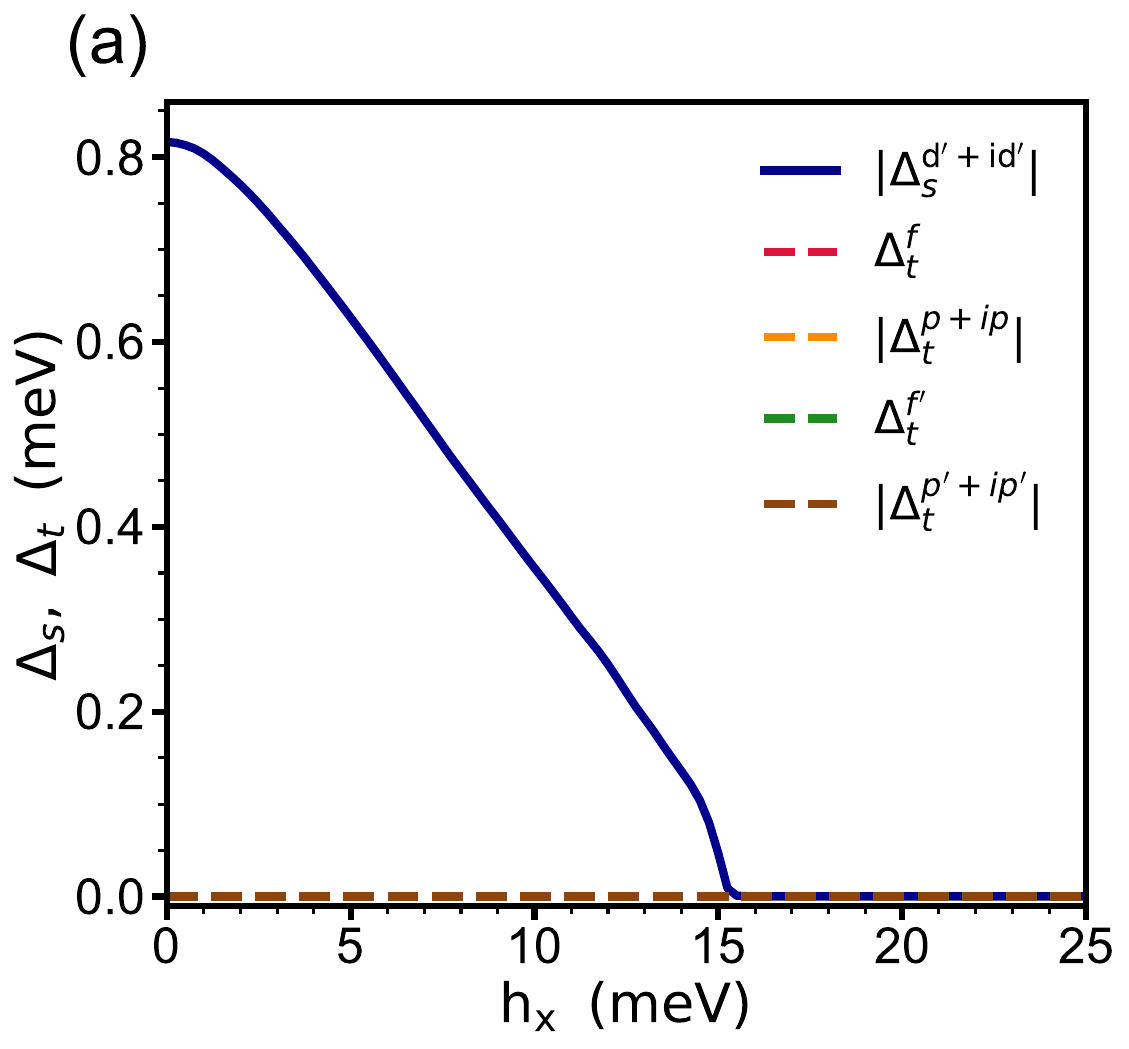}
		\end{minipage}
		\hfill
		\begin{minipage}[t]{0.28\linewidth}
			\centering
			\includegraphics[width=\linewidth]
			{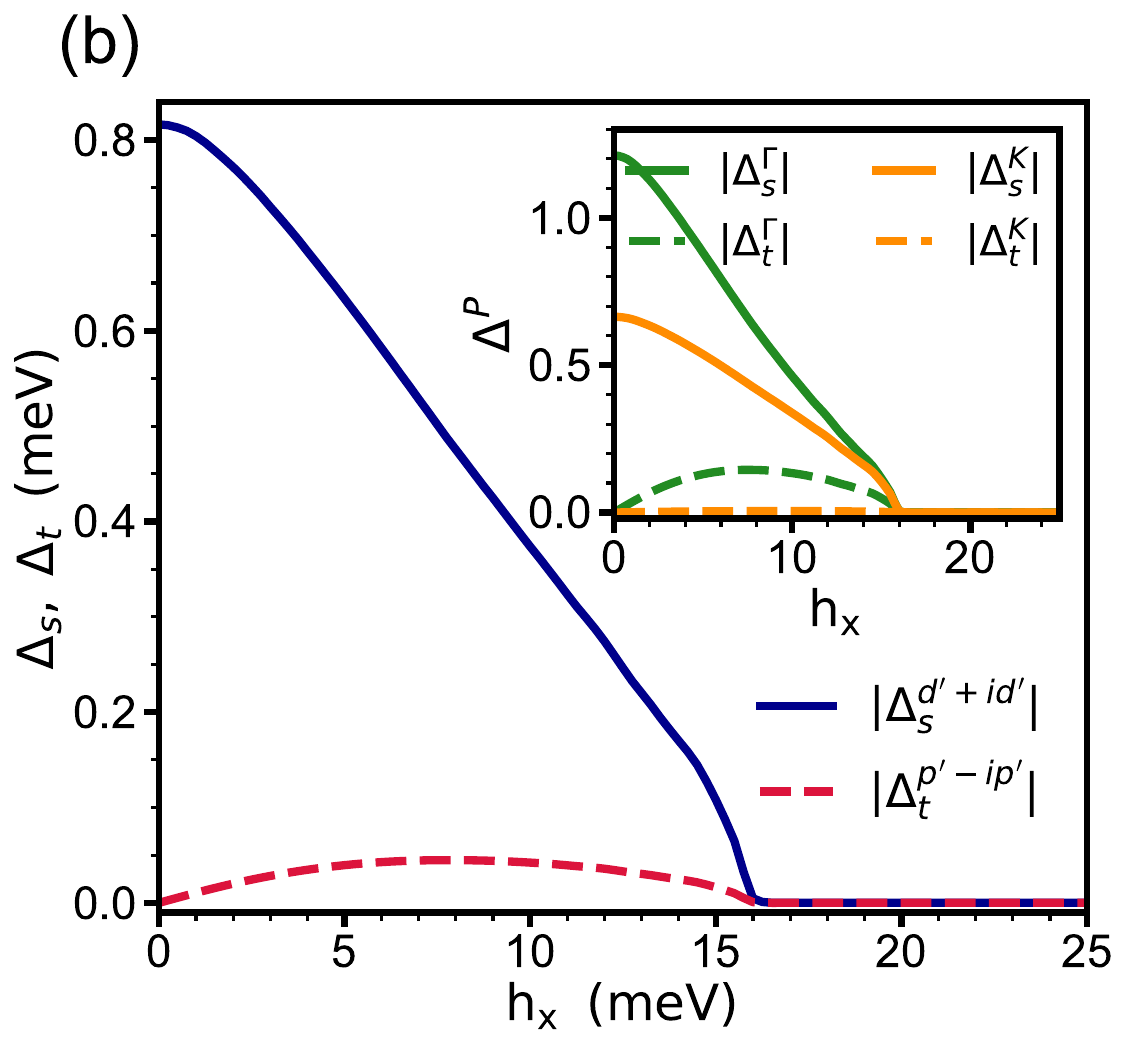}
		\end{minipage}
		\hfill
		\begin{minipage}[t]{0.28\linewidth}
			\centering
			\includegraphics[width=\linewidth]
			{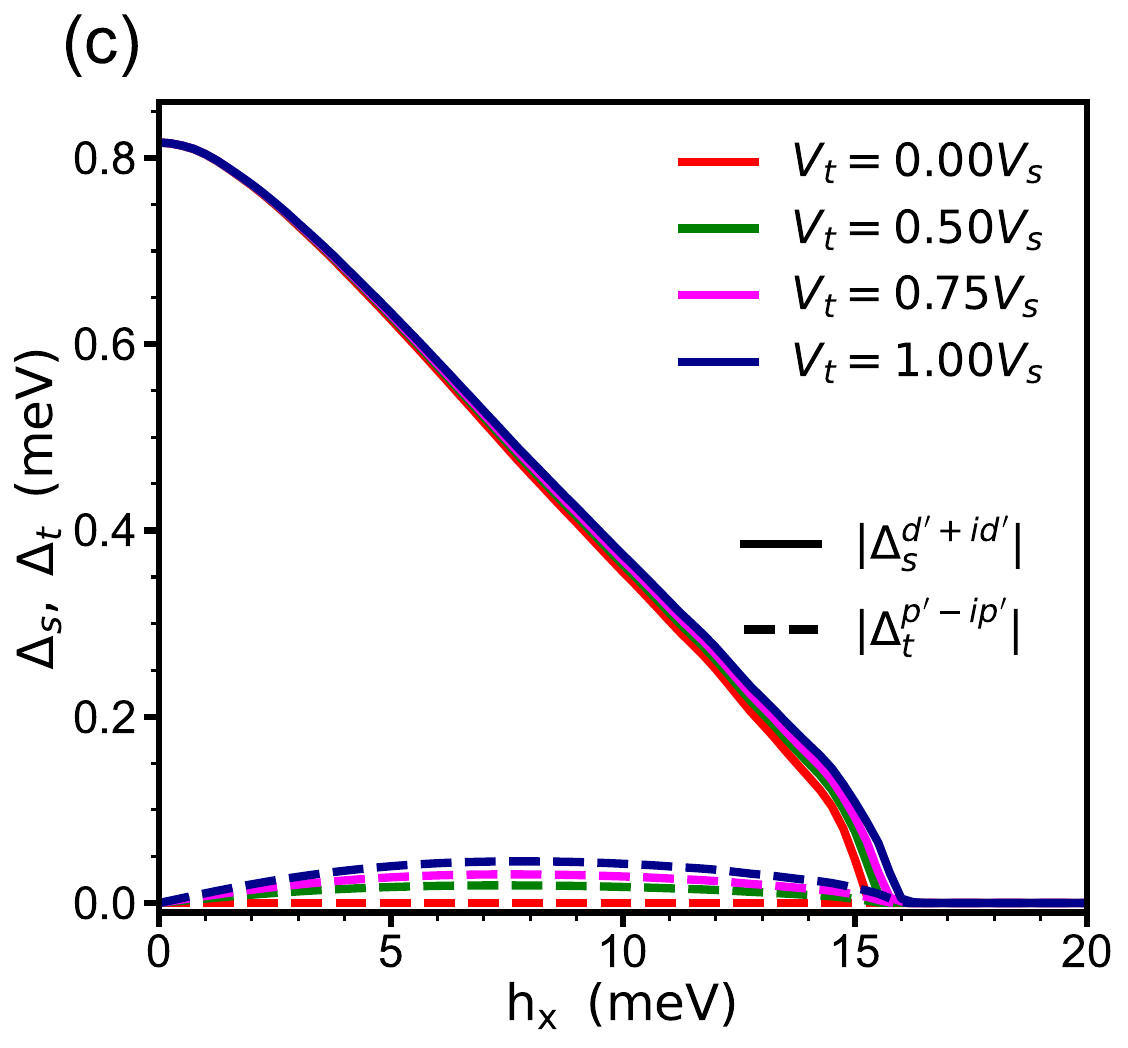}
		\end{minipage}
		
		\caption{
			Field dependence of the self-consistent equal-spin triplet order parameters for the candidate odd-parity channels of the NNN chiral $d^\prime+id^\prime$ state. The channels shown in (a) remain zero, while the NNN $p^\prime-ip^\prime$ component develops a finite order parameter in (b), with the inset showing the corresponding $\Gamma$- and $K$-resolved singlet and triplet contributions. The dependencies of the order parameter on $V_t$ is shown in (c).
		}
		
		\label{fig:NNN_chiral_d_triplet_channels}
	\end{figure*}

	The NNN chiral $d'+id'$ singlet state is further examined for all
	candidate equal-spin triplet channels induced by the in-plane field in
	Fig.~\ref{fig:NNN_chiral_d_triplet_channels}. The NN
	$p-ip$ component discussed in the main text
	[Fig.~\ref{fig:NNN_triplet}(a)] is examined together with the remaining
	candidate odd-parity channels. Figure~\ref{fig:NNN_chiral_d_triplet_channels}(a)
	shows that the NN $f$-wave and $p+ip$, NNN $f^\prime$-wave, and NNN
	$p^\prime+ip^\prime$ components remain zero throughout the field range.
	
%
%
%
%
	
	Figure~\ref{fig:NNN_chiral_d_triplet_channels}(b) shows the field
	dependence of the NNN $p^\prime-ip^\prime$ component together with the
	parent singlet order parameter. The inset shows their $\Gamma$- and
	$K$-resolved contributions, indicating that the induced triplet
	component is associated with the $\Gamma$ pocket, while the valley contribution at
	$K$  is absent.
	
	The dependence on the strength of interaction in  triplet pairing channel, $V_t$, on the order parameters is shown in
	Fig.~\ref{fig:NNN_chiral_d_triplet_channels}(c). Increasing $V_t$
	enhances the induced $p^\prime-ip^\prime$ component, but produces only
	a modest increase in the critical field. This dependence is considerably
	weaker than that of the NN $p-ip$ state shown in
	Fig.~\ref{fig:NNN_triplet}(b). Thus, although the NNN $d'+id'$ state
	selects the opposite-chirality $p^\prime-ip^\prime$ triplet component,
	its magnitude and influence on the critical field are weaker than for
	the corresponding NN pairing.
	
	
	\bibliography{references}
	
\end{document}